\documentclass[11pt,a4paper]{article}
\pdfoutput=1

\usepackage{amsmath,amssymb,tikz}
\usepackage{graphicx}
 \allowdisplaybreaks
\def\be{\begin{equation}}
\def\ee{\end{equation}}
\def\ba#1\ea{\begin{align}#1\end{align}}
\def\bg#1\eg{\begin{gather}#1\end{gather}}
\def\bm#1\em{\begin{multline}#1\end{multline}}
\def\bmd#1\emd{\begin{multlined}#1\end{multlined}}

\def\d{\delta}

\def\e{\epsilon}

\def\g{\gamma}

\def\s{\sigma}

\def\la{\label}

\def\nn{\nonumber}

\def\({\left(}
\def\){\right)}
\def\[{\left[}
\def\]{\right]}

\def\cO{{\mathcal O}}

\def \be {\begin{equation}}
\def \ee {\end{equation}}
\def \ba {\begin{array}}
\def \ea {\end{array}}
\def \bea{\begin{eqnarray}}
\def \eea{\end{eqnarray}}
\def \nn {\nonumber}

\def \g {\gamma}

\def \d {\delta}

\def \e {\epsilon}

\def \s {\sigma}

\def \th {\theta}

\def \la {\leftarrow}
\def \ra {\rightarrow}

\def\bea{\begin{eqnarray}}
\def\eea{\end{eqnarray}}
\newcommand{\eq}[1]{(\ref{#1})}
\newcommand{\bit}{\begin{itemize}}  \newcommand{\eit}{\end{itemize}}
\newcommand{\ben}{\begin{enumerate}}  \newcommand{\een}{\end{enumerate}}

\def\la{\langle}
\def\ra{\rangle}

\def\cA{{\cal A}}

  \def\cO{{\cal O}}

 \def\del{\partial}

\long\def\symbolfootnote[#1]#2{\begingroup%
\def\thefootnote{\fnsymbol{footnote}}\footnote[#1]{#2}\endgroup}

\newcommand{\nthu}{{\it Department of Physics, National Tsing-Hua
  University,
Hsinchu 30013, Taiwan}}

\newcommand{\ncts}{{\it Physics Division, National Center for Theoretical Sciences,
National Tsing-Hua University, Hsinchu 30013, Taiwan}}

\newcommand{\sysu}{{\it School of Physics and Astronomy, Sun Yat-Sen University, 2 Daxue Road, Zhuhai 519082, China}}

\begin{document}
\thispagestyle{empty}
\hfill{NCTS-TH/2007} 
\begin{center}

~\vspace{20pt}

{\Large\bf Weyl Anomaly induced Fermi Condensation and Holography}

\vspace{25pt}

Chong-Sun Chu ${}^{2,3}$\symbolfootnote[1]{Email:~\sf
  cschu@phys.nthu.edu.tw}
, Rong-Xin Miao ${}^1$\symbolfootnote[2]{Email:~\sf
  miaorx@mail.sysu.edu.cn}

\vspace{10pt}${{}^{1}}$\sysu \footnote{All the Institutes of authors
  contribute equally to this work, the order of Institutes is adjusted
  for the assessment policy of SYSU.}

\vspace{10pt}${{}^{2}}$\ncts

\vspace{10pt}${{}^{3}}$\nthu

\vspace{2cm}

\begin{abstract}
Recently it is found that, due to Weyl anomaly, a background scalar
field induces a non-trivial
Fermi condensation for theories with Yukawa couplings.  For
simplicity,
the paper consider only
scalar type Yukawa coupling and, in the BCFT case,  
only for a specific boundary condition. In these cases, the
Weyl anomaly takes
on a simple special form. 
In this paper, we generalize
the results to
more
general situations.
First, we obtain general expressions of Weyl anomaly
due to a background scalar and pseudo scalar field in
general 4d BCFTs.  Then, we derive the
general
form of Fermi condensation from the Weyl anomaly.  It is remarkable that,
in general, Fermi condensation is non-zero even if there was not a non-vanishing
scalar field background.
Finally, we verify our results with free BCFT
with Yukawa coupling to scalar and pseudo-scalar background potential with  
general chiral bag boundary condition
and
with holographic BCFT. In particular, we obtain the shape and curvature
dependence of the Fermi condensate from the  holographic one point function.
\end{abstract}

\end{center}

\newpage
\setcounter{footnote}{0}
\setcounter{page}{1}

\tableofcontents

\section{Introduction}

Similar to Bose Einstein condensation, Fermi condensation is an
interesting quantum phenomena, which has wide a range of
applications. The famous examples include the Cooper pair in BCS
theory of superconductivity, which is the bound state of a pair of
electrons in a metal with opposite spins. The chiral condensate of
massless fermions is another example of Fermi condensation. In QCD the
chiral condensate is an order parameter of transitions between
different phases of quark matter in the massless limit. The condensation
of fermionic atoms has been observed in experiment
\cite{fermicondensation}.

Recently, it is found that Weyl anomaly can induce Fermi condensation
for theories with Yukawa couplings \cite{Chu:2020mwx}, when a
background scalar is turned on.  The mechanism is similar to the those
of Weyl anomaly induced Casimir effect \cite{Miao:2017aba} and current
\cite{Chernodub:2016lbo,Chernodub:2017jcp,Chu:2018ksb}.
For simplicity, \cite{Chu:2020mwx} discusses 
only the free Dirac fermion theory with the action
\begin{eqnarray} \label{Fermiaction}
  I=\int_M \sqrt{|g|} \left( \bar{\psi}i \gamma^{i}\nabla_{i}\psi+
  \phi \bar{\psi} \psi\right),
\end{eqnarray}
where $\bar{\psi}=\psi^\dagger \gamma^0$ and $\phi$ is a background scalar field.
We take signature $(1,-1,-1,-1)$ in this paper. The gamma matrix obeys
\begin{eqnarray}\label{notation2}
\{ \gamma^{i},  \gamma^{j} \}=2\ g^{ij}.
\end{eqnarray}
Imposing the following bag boundary condition (BC)
\cite{Chodos:1974je,Chodos:1974pn,Vassilevich:2003xt}
\begin{eqnarray} \label{bagBCspecial}
 (1\pm \gamma_5 \gamma^i n_i)\psi|_{\partial M}=0
\end{eqnarray}
and applying the heat kernel expansion \cite{Vassilevich:2003xt},
\cite{Chu:2020mwx} gets Weyl anomaly at one loop
\begin{eqnarray} \label{Fermianomaly}
  \mathcal{A}=\frac{1}{8\pi^2}\left(\int_M \sqrt{|g|}
  \Big[-(\nabla \phi)^2+\frac{R\phi^2}{6}+\phi^4\Big]
  +\int_{\partial M}\sqrt{|h|} \frac{k \phi^2}{3} \right).
\end{eqnarray}
Here $k=\nabla_{i} n^{i}$ and 
$n^{i}$ is the outward-pointing normal vector. 
From the action (\ref{Fermiaction}), it is clear that the Fermi condensation
is given by the renormalization expectation value of the
scalar operator $\cO := \bar{\psi} \psi$,
\begin{eqnarray} \label{Fermicondensation}
\la \bar{\psi} \psi \ra = \frac{1}{\sqrt{|g|}}\frac{\delta I_{\rm eff}}{\delta \phi},
\end{eqnarray}
where $I_{\rm eff} $ is the effective action of fermions.
For a flat half space $x\ge 0$,  
it is remarkable that the 
Fermi condensation (\ref{Fermicondensation}) can
be derived from Weyl anomaly  (\ref{Fermianomaly}) as \cite{Chu:2020mwx}
\begin{eqnarray}
  \label{FermicondensationIIntro}
\la \bar{\psi} \psi \ra= -\frac{1}{4\pi^2}\frac{n^i\nabla_{i}\phi
  +\frac{1}{3}k\phi}{x} +O(\ln x), \quad x \sim 0,
\end{eqnarray}
where we have used $n^i\nabla_{i}\phi=-\partial_x \phi$ since $n^{i}=(0,-1,0,0)$.

In this paper, we generalize the work of \cite{Chu:2020mwx} to more general
class of boundary conditions in four dimensional CFT/BCFT
\cite{Cardy:2004hm,McAvity:1993ue}. We show that, by imposing the Wess-Zumino
consistency condition, one can obtain the general expression of Weyl anomaly
due to a background scalar field (or pseudoscalar field) $\phi$
\footnote{The same results apply for a pseudoscalar field. In the rest of the
  paper, unless otherwise stated,
  we will refer to both a scalar and a pseudoscalar simply as a scalar
  without specifying its parity.}.
Compared with (\ref{Fermianomaly}), generally more
boundary terms are allowed to appear. This is one of the main
results of this paper. We then show that the
presence of the Weyl anomaly implies that the scalar operator defined by
\be \label{classical-O}
\cO := \frac{1}{\sqrt{|g|}}\frac{\delta I}{\delta \phi}
\ee
obtains a nontrivial
expectation value near the boundary. Generally new contributions
that are independent of the background scalar field can arise.
We show that this also occur in conformally flat spacetime without
boundaries.
This is another interesting result of this paper. 
Finally, we verify our results with the Yukawa theory of fermions
coupled to a background scalar or  pseudoscalar field with general
BCs. We do the same for the holographic BCFT and
we obtain, in particular, the shape and
curvature dependence of the one point function of the dual scalar
operator in strongly coupled CFT. This is an interesting quantity and we
expect it to have non-trivial implications on the phase structure of the theory. 

The paper is organized as follows.  In section  2, we obtain the general
expressions of Weyl anomaly for 4d BCFTs with
a general shape of boundary in a curved spacetime, and in
the presence of a background scalar
field. In section  3, we show that the Weyl anomaly induces
a condensation for the corresponding scalar operator $\cO$
in a  BCFT near the boundary or in a CFT in a
conformally flat spacetime without
boundaries. In section  4, we consider the Yukawa theory with general
BCs and verify the anomalous Fermi condensation near the boundary.  In
section  5, we study the holographic one point function near the boundary
of BCFT and verify that it takes the expected form as derived in section  3. In
section  6, we give a holographic proof of the Weyl anomaly induced one-point
function in
conformally flat spacetime without boundaries. Finally, we conclude in
section  7.

Conventions. People in the fields of quantum field theory and gravity
theory usually use different signature of the metric
\cite{Parker:2009uva,Carroll:2004st}.  For the convenience of the
reader, we take signature $(1,-1,-1,-1)$ in
section 1- section 4 for the
field-theoretical discussions, while signature $(1,1,1,1)$ or
$(-1,1,1,1)$ in section 5 and section 6 for the holographic study.  In
signature $(1,-1,-1,-1)$ \cite{Parker:2009uva}, $R^i_{\ jkl}=
\partial_l \Gamma^i_{jk} - \partial_k \Gamma^i_{jl}
+\Gamma^i_{lm}\Gamma^m_{jk} -\Gamma^i_{km}\Gamma^m_{jl}$,
$R_{ij}=R^l_{\ i l j}$, $R=g^{ij} R_{ij}$, $k_{ij}=h^k_i h^l_j\nabla_k
n_{l}$, $k=h^{ij} k_{ij}=\nabla_i n^i$ where $n^i$ is the normal
vector given by
$n^{i}=-n_i=(0,-1,0,0)$ in a flat half space $x\ge 0$.
While in signature $(1,1,1,1)$ and $(-1,1,1,1)$ \cite{Carroll:2004st},
$R^i_{\ jkl}= \partial_k \Gamma^i_{jl} - \partial_l \Gamma^i_{jk}
+\Gamma^i_{km}\Gamma^m_{jl} -\Gamma^i_{lm}\Gamma^m_{jk}$,
$R_{ij}=R^l_{\ i l j}$, $R=g^{ij} R_{ij}$, $k_{ij}=h^k_i h^l_j\nabla_k
n_{l}$, $k=h^{ij} k_{ij}$ where $n^i$ is the outward-pointing normal
vector.  Note that Fermi condensation $\la \bar{\psi} \psi \ra$,
stress tensor $T_{ij}$, Ricci scalar $R$, normal vector $n^i$ and the
trace of extrinsic curvature $k$ are the same, while $R^i_{\ jkl},
R_{ij}, k_{ij}, g_{ij}, n_i$ and, in particular, the Weyl anomaly
$\la T^i_{ i} \ra$, differ by a minus sign in different signatures.  Note
that $R$ and $k$ agree with those of \cite{Vassilevich:2003xt} in both
signatures.

\section{Weyl Anomaly due to Scalar Background}

Let $\phi$ be a scalar field or pseudo-scalar field with
dimension one, which we will consider as a background.
Similar to the background gravitational field and gauge
field, it leads to Weyl anomaly \cite{Duff:1993wm}. For a CFT/BCFT,
the Weyl anomaly should be Weyl invariant and obey the Wess-Zumino
consistency condition \cite{Wess:1971yu}
\begin{eqnarray}\label{WZconsistency}
[\delta_{\sigma_1},\delta_{\sigma_2}] \mathcal{A}=0.
\end{eqnarray}
Imposing the above conditions, we obtain the general expressions of
Weyl anomaly due to a background field $\phi$:
\begin{eqnarray}\label{anomaly}
\mathcal{A}= a_1 \mathcal{A}_1+ a_2 \mathcal{A}_2+ \sum_{n=1}^{4} b_n
\mathcal{B}_n,
 \end{eqnarray}
where
$\mathcal{A}_n, \mathcal{B}_m$ are given by
\begin{eqnarray}\label{A1}
&& \mathcal{A}_1= \int_M \sqrt{|g|} [-(\nabla
    \phi)^2+\frac{1}{6}R\phi^2 ]+\int_{\partial M}\sqrt{|h|}
  \frac{1}{3}k\phi^2, \\ \label{A2} 
  && \mathcal{A}_2= \int_M
  \sqrt{|g|} \phi^4, \\ \label{B1} 
  && \mathcal{B}_1= \int_{\partial
    M}\sqrt{|h|} \phi^3,\\ \label{B2} 
  && \mathcal{B}_2= \int_{\partial
    M}\sqrt{|h|}[ k \phi^2+3\phi n^{i}\nabla_i \phi ],\\ \label{B3} &&
  \mathcal{B}_3= \int_{\partial M}\sqrt{|h|}[ R \phi +6 \Box \phi
  ],\\ \label{B4} 
  && \mathcal{B}_4= \int_{\partial M}\sqrt{|h|}[ {\text{Tr}} \bar{k}^2 \phi ]
\end{eqnarray}
and $a_n, b_m$ are the corresponding bulk and boundary central charges. 
 Here $g_{ij}, R, \nabla_i, \Box$ are metrics, Ricci scalar, covariant
 derivatives and D'Alembert operator defined in the bulk $M$, $h_{ab}$
 is the induced metric on the boundary $\partial M$, 
 $n^{i}$ is the
 outpointing normal vector given by
 $n^i=-n_i=(0,-1,0,0)$
 in a flat half space, $k_{ab}=h_a^{\ i}h_b^{\ j}\nabla_i n_j$ is
 the extrinsic curvature and $\bar{k}_{ab}=k_{ab}-\frac{1}{3}kh_{ab} $
 is its traceless part.
 
 Some comments are in order. {\tt 1.} The bulk central charges $a_n$ are
 independent of boundary conditions, while the boundary central
 charges $b_m$ depend on boundary conditions.
 {\tt 2.} Second, as mentioned
 above, the Weyl anomaly (\ref{anomaly}) obeys the Wess-Zumino
 consistency (\ref{WZconsistency}).
 {\tt 3.} We consider only integer
 powers of $\phi$ and ignore terms including $\phi^{1/2},
 \phi^{1/3},...$. If such terms are allowed, we could construct
 scalar-invariant terms such as
 \begin{eqnarray}\label{possible}
\int_{\partial M}\sqrt{|h|}[-(D \phi^{\frac{1}{2}} )^2+\frac{1}{8}\mathcal{R}\phi ],
 \end{eqnarray}
 where $D_a, \mathcal{R}$ are covariant derivatives and Ricci scalar
 on the boundary $\partial M$, respectively. However, since $(D
 \phi^{\frac{1}{2}} )^2=\frac{1}{4}(D \phi)^2/ \phi$ is not
 well-defined on points with $\phi=0$ but $D\phi \ne 0$, we rule out
 such possible contributions to Weyl anomaly.
 {\tt 4.} We focus on
 CFT/BCFT in this paper. For general QFT, non-scale-invariant terms
 are allowed in Weyl anomaly.
 {\tt 5.} We can rewrite $\mathcal{B}_3$
 into more convenient form for the purpose to derive Fermi
 condensation
 \begin{eqnarray}\label{B3a}
   \mathcal{B}_3= \int_{\partial M}\sqrt{|h|}[ R \phi -6 n^i n^j \nabla_i \nabla_j
     \phi-6 k n^i \nabla_i \phi+6 D^aD_a \phi],
 \end{eqnarray}
where the total derivative term $ D^aD_a \phi$ can be dropped since
$\partial M$ is closed, i.e., $\partial (\partial M)=0$. In the next
section, we shall show that $ \mathcal{B}_3$ is related to the leading
term of Fermi condensation near the boundary.

\section{Anomalous Condensation}

In this section, we show that in
four dimensional spacetimes with and without boundaries,
the operator $\cO$ that couples to the scalar field $\phi$ obtains
a non-trivial expectation value due to the Weyl anomaly \eq{anomaly}. For
simplicity, we focus on the case of CFT/BCFT below.
For the theory of Dirac fermions  with
Yukawa coupling to a background scalar field $\phi$,
$\cO =\bar{\psi} \psi$ and 
the expectation value  $\la \cO\ra$ gives Fermi condensation.

\subsection{Spacetime with Boundary}

Let us first investigate the case with boundaries. Since the mass
dimension of scalar operator $O$ is three, its expectation value takes
the asymptotic form \cite{Deutsch:1978sc}
\begin{eqnarray} \label{FermicondensationI0}
  \la \cO \ra
  =\frac{O_0}{x^3}+\frac{O_1}{x^2}+\frac{O_2}{x}+O(x^0,\ln
x)
\end{eqnarray}
near the boundary. Here $x$ is the proper distance from the boundary,
$O_n$ have mass dimension $n$ and depend on only the background
geometry and background scalar. Below we will derive exact expressions
of $O_n$ from the Weyl anomaly.

One way to see that the coefficients $O_n$ are directly connected
with the Weyl anomaly is by noticing that
the one point function \eq{FermicondensationI0}
can be understood as a well-defined distribution  \cite{McAvity:1990we,Petkou}
if the inverse powers of $x$ are accompanied by logarithmically
divergent contact terms $\ln x\; \partial^{n}_x \delta(x)$.
Such contract terms determine the scale variation of $\la O \ra$ and hence
the coefficients $O_n$ of (\ref{FermicondensationI0}) are in
fact determiend by the central charges of Weyl anomaly
\footnote{We thank the referee
  for emphasising this point to us.}.  From this point of view,
it is clear that the coefficient $O_0$ is completely determined by an
anomaly coefficient instead of other non-anomalous data.

In this paper, we use an alternative method to derive $O_n$ from the
Weyl anomaly.
Using the fact that 
the Weyl anomaly is related to the UV Logarithmic divergent
term of the effective action, one can \cite{Miao:2017aba,Chu:2018ksb}
establish the relation
\begin{eqnarray} \label{key0}
(\delta \mathcal{A})_{\partial M}= (\delta
  I_{\rm eff})_{\log \epsilon}=\left(\int_M \sqrt{|g|}
  (\frac{1}{2}\la T_{ij} \ra \delta g^{ij}+
  \la J_i\ra \d A^i + \la \cO \ra \delta
  \phi)\right)_{\log\epsilon}
\end{eqnarray}
which relates directly the variation of the Weyl anomaly with
a corresponding one-point function.
Here a regulator $x \geq \e$ to the boundary has been introduced in
the integral on the RHS of \eq{key0}
and the symbol $(\; )_{\log \e}$ denotes
the coefficient of the $\log \e$ term.
The first equation of \eq{key0} is due
to the definition of Weyl anomaly, and the second equation of
\eq{key0} is just the definition of one point functions.  For our
purpose, we will turn on only the variation of scalar and focus on
\begin{eqnarray} \label{key}
  (\delta_{\phi} \mathcal{A})_{\partial M}=\left(\int_M \sqrt{|g|} \la \cO\ra
  \delta \phi \right)_{\log \epsilon}.
\end{eqnarray}
The variations $\d g^{ij}, \d A^i$, $\d \phi$ are independent. Previously
the one-point functions $\la T_{ij} \ra, \la J_i\ra$
have been studied.
In this paper we will consider the scalar variation $\d \phi$ and derive the
one-point function $\la \cO \ra$ from the Weyl anomaly.

To proceed, let us consider the metric written in the Gauss normal coordinates
\begin{eqnarray} \label{Gausscoordinate}
ds^2=-dx^2+ \left(h_{ab}(y)-2x k_{ab}(y)+ x^2 q_{ab}(y)+O(x^3) \right)dy^ady^b 
\end{eqnarray}
and expand the scalar near the boundary as 
\begin{eqnarray} \label{scalar0}
\phi(x,y)=\phi_0(y)+ x \phi_1(y)+ \frac{x^2 }{2}\phi_2(y)+O(x^3)
\end{eqnarray}
where $n^i=-n_i=(0,-1,0,0)$ and $\phi_m$ are independent variables.  From
(\ref{anomaly}), we get the LHS of (\ref{key})
\begin{eqnarray} \label{keyLHS}
&&\int_{\partial M}\sqrt{|h|} \left(-6 b_3 \delta \phi_2 + \left(6 b_3
  k-3 b_2 \phi_0\right) \delta
  \phi_1\right)\nonumber\\ &+&\int_{\partial M}\sqrt{|h|}
  \left(\frac{2}{3} a_1 k \phi _0-2 a_1 \phi _1+2 b_2 k \phi _0+b_3
  R+b_4 \text{Tr}\bar{k}^2+3 b_1 \phi _0^2-3 b_2 \phi _1\right)\delta
  \phi_0
\end{eqnarray}
 Next, we substitute (\ref{FermicondensationI0}) into the RHS of
 (\ref{key}), integrate over $x$ and select the logarithmic divergent
 term, we obtain
\begin{eqnarray} \label{keyRHS}
&&\int_{\partial M}\sqrt{|h|} \left( -\frac{O_0}{2} \delta \phi_2 -
  \left(O_1-k O _0\right) \delta
  \phi_1\right)\nonumber\\ &+&\int_{\partial M}\sqrt{|h|} \left(
  -\frac{1}{2} O _0 \left(k^2+q-2 \text{Tr}k^2\right)+k O_1-O_2\right)\delta \phi_0,
\end{eqnarray}
where we have used $\sqrt{|g|}=\sqrt{|h|}\left(1-k x+\frac{1}{2}
\left(k^2+q-2 \text{Tr} k^2\right)x^2+O(x^3)\right)$ and
$q=h^{ab}q_{ab}$ in the above calculations. Comparing (\ref{keyLHS})
and (\ref{keyRHS}), we can solve
\begin{eqnarray} \label{results1}
&&O_0=12 b_3, \ O_1=3 \left(2 b_3 k+b_2 \phi _0\right),\nonumber\\
&& O_2=-2a_1(\frac{1}{3}  k \phi_0-  \phi_1)-3 b_1 \phi_0^2
  +b_2( k \phi_0+3  \phi_1)-b_3 (6 q+R-12 \text{Tr}k^2)-b_4 \text{Tr}\bar{k}^2.
  \;\;\;\;
\end{eqnarray}
From \eq{Fermicondensation}, \eq{FermicondensationI0} and \eq{results1},
we finally obtain one of our main results for the expectation value of the
Fermi condensation near the boundary:
\begin{eqnarray} \label{FermicondensationGoodI}
\la\bar{\psi} \psi\ra&=&\frac{12 b_3}{x^3}+\frac{6 b_3 k+3b_2 \phi}{x^2}\nonumber\\
&+&\frac{ -2a_1(n^i\nabla_i+\frac{1}{3}  k)\phi
  -3 b_1 \phi^2 +b_2 k \phi
  -b_3 (R+6 R_{nn}-6\text{Tr}k^2) -b_4 \text{Tr}\bar{k}^2}{x}\nonumber\\
&+&O(x^0,\ln x),
\end{eqnarray}
where $\phi = \phi(x) = \phi_0 + x \phi_1 + \cdots$ in the above expression.
Above we have rewritten $O_n$ into covariant expressions and have used
$R_{nn}=q-\text{Tr}k^2$ in Gauss normal coordinates. 

Let us make some comments. 
{\tt 1}. (\ref{FermicondensationGoodI}) shows that the leading terms
of Fermi condensation near the boundary are completely fixed by
central charges of Weyl anomaly. In general, the boundary central
charge depends on choices of boundary conditions, so does the Fermi
condensation (\ref{FermicondensationGoodI}). {\tt 2}. Similar to the
case of current and stress tensor \cite{Chu:2018ksb,Miao:2017aba},
there are boundary contributions to the Fermi condensation, which can
cancel the ``bulk divergence" and make finite the total Fermi
condensation.  {\tt 3}. (\ref{FermicondensationGoodI}) works for
general 4d BCFTs.  For non-BCFTs, there are corrections to Weyl
anomaly and thus corresponding corrections to Fermi condensation
(\ref{FermicondensationGoodI}) .  {\tt 4}.
(\ref{FermicondensationGoodI}) agrees with the results of
the
free theory with $b_i=0$ \cite{Chu:2020mwx}
\begin{eqnarray} \label{FermicondensationGoodI2}
\la\bar{\psi} \psi\ra =\frac{- 2a_1(n^i \nabla_i+\frac{1}{3}
  k)\phi}{x}+O(x^0,\ln x)
\end{eqnarray}
Note that $\nabla_n$ of \cite{Chu:2020mwx} denotes $\nabla_x$, so it
is given by $-n^i \nabla_i$ in this paper.
    {\tt 5}.
    In general in a curved spacetime and for curved boundary,
  the Fermi condensation (\ref{FermicondensationGoodI}) is
  non-vanishing even without a background scalar
  \begin{eqnarray} \label{FermicondensationGoodI3}
    \la\bar{\psi} \psi\ra_{\phi=0}=
    \frac{12 b_3}{x^3}+\frac{6 b_3 k}{x^2}-\frac{b_3 (R+6 R_{nn}-6\text{Tr}k^2)
  +b_4 \text{Tr}\bar{k}^2}{x}+O(x^0,\ln x).
\end{eqnarray}
This generalize the result of \cite{Chu:2020mwx}.

\subsection{Conformally Flat Spacetime without Boundary}

Let us next turn to discuss the case without boundaries.
For simplicity, we focus on conformally flat spacetime.
Let us start by deriving the anomalous
transformation rule for the condensate. 
Consider a theory with metric and scalar field given by
$(g_{ij},\phi)$. Due to the anomaly, the renormalized effective
action $I_{\rm eff}$ is not invariant under the Weyl transformation.
Consider the Weyl transformation
\be
g_{ij} \to g_{ij}'= e^{-2\s} g_{ij}, \quad \phi \to \phi' = e^\s \phi,
\ee
for arbitrary finite $\s(x)$, we have generally
\be
\frac{\d}{\d\s} I_{\rm eff}(e^{-2\s} g_{ij}, e^{\s}\phi) =
\cA (e^{-2\s} g_{ij}, e^{\s}\phi).
\ee
This can be integrated to give the effective
action \cite{Wess:1971yu,Cappelli:1988vw,Schwimmer:2010za}. Using the fact that
the anomaly (\ref{anomaly}) is Weyl invariant up to a surface term:
\be
\cA(e^{-2\s} g_{ij}, e^{\s} \phi)
= \cA(g_{ij}, \phi) + a_1 \int_M
\del_i(\sqrt{-g} \phi^2 g^{ij} \del_j \s),
\ee
we obtain the transformation rule for the effective action:
\bea\label{Ieff-t}
&& I_{\rm eff}(e^{-2\s} g_{ij}, e^{\s}\phi)=  I_{\rm eff}(g_{ij}, \phi) \nn  \\
&& + a_1\int_M \sqrt{|g|}
\left[\left( -(\nabla \phi)^2+\frac{R\phi^2}{6}\right)\s
  +\frac{\phi^2}{2}(\nabla \sigma)^2\right]
+ \frac{a_1}{3}\int_{\partial M}\sqrt{|h|} k \phi^2 \sigma \nn\\
&&+ a_2\int_M \sqrt{g} \phi^4 \sigma + \sum_{n=1}^4 b_n \mathcal{B}_n \sigma.
\end{eqnarray}
One can check that the effective action satisfies Wess-Zumino
consistency $[\delta_{\sigma_1},\delta_{\sigma_2}] I_{\rm eff}=0$. This is
a test of our results.  Using \eq{Ieff-t},
we obtain finally the anomalous transformation rule for the 
condensate \eq{Fermicondensation} under Weyl transformation
$g_{ij} \to g'_{ij} = e^{-2\sigma} g_{ij}$, $\phi \to \phi' = e^\s \phi$,
\bea \label{FermicondensationI3}
\la O \ra =
- 2a_1 \nabla(\sigma\nabla\phi)
-(\frac{a_1}{3}\phi R+ 4 a_2 \phi^3) \sigma
-a_1 \phi (\nabla\sigma)^2, 
  \eea
plus the term  $e^{-3\s} \la O\ra'$
and some boundary terms which we drop in spacetime without boundaries.
  Here $\la O \ra$ (resp.  $\la O \ra'$)
  denotes the vev of the condensate of
  the theory \eq{action} in the background spacetime $g_{ij}$
  (resp. $g'_{ij}$).
Taking $g'_{ij}$ to be the flat spacetime metric and the fact that the
Fermi condensation vanishes in flat spacetime,
we finally obtain \eq{FermicondensationI3} as the
Fermi condensate in conformally flat spacetime 
\be\label{cf-metric}
ds^2 = e^{2 \s} \eta_{ij}dx^i dx^j.
\ee
For Dirac fermions with Yukawa coupling, we have $O= \bar{\psi} \psi$,
$a_1 = 1/(8\pi^2)$
and \eq{FermicondensationI3} reproduces the result of \cite{Chu:2020mwx}.

\section{Yukawa Coupled Fermions}

In this section, we investigate the anomalous Fermi condensation for the Yukawa
coupled Dirac theory \eq{Fermiaction} with more general BCs.
We will derive the general expression \eq{anomaly} for the
Weyl anomaly
and also the corresponding Fermi condensate.

The BCs of Dirac fields should make zero the normal current on the
boundary.  According to \cite{McAvity:1993ue}, the general BCs take
the form
\begin{eqnarray}\label{generalBBC}
\Pi_{-} \psi|_{\partial M}=0,
\end{eqnarray}
where $\Pi_{\pm}=(1\pm \chi)/2$ are projection operators and $\chi$
satisfy \cite{McAvity:1993ue}
\begin{eqnarray}\label{gamma}
\chi \gamma^n=-\gamma^n \bar{\chi}, \ \chi \gamma^a=\gamma^a
\bar{\chi}, \ \chi^2=\bar{\chi}^2=1.
\end{eqnarray}
Here $\bar{\chi}=\gamma^0 \chi^+ \gamma^0$ and $n$ ($a$) denote the
normal (tangent) directions.  Without loss of generality, we choose
\begin{eqnarray}\label{chi}
 \chi=-i e^{i \theta \gamma_5 } \gamma^{i}n_{i},
 \end{eqnarray}
 which defines the so-called chiral bag boundary condition
\begin{eqnarray}\label{chiralBC}
(1+ i e^{i \theta \gamma_5 } \gamma^{i}n_{i}) \psi|_{\partial M}=0.
\end{eqnarray}
Here $\theta$ is a constant and 
$n_{i}$ is the normal vector given by
$(0,1,0,0)$ in a flat half space.
Note that the BC (\ref{chiralBC}) reduces to the usual bag BC
$(1 \pm i \gamma^{i}n_{i}) \psi|_{\partial M}=0$ for
$\theta=0,\pi$. And it reduces to the BC (\ref{bagBCspecial}) studied in
\cite{Chu:2020mwx} when $\theta=\pm \frac{\pi}{2}$.

From the BC \eq{chiralBC} and EOM $(i \gamma^i
\nabla_i+\phi)\psi=0$, one can derive that
\begin{eqnarray}\label{PpBBC}
\left( -n^i\nabla_i +S \right)\Pi_{+} \psi|_{\partial M}=0,
\end{eqnarray}
where
\begin{eqnarray}\label{S}
S=- (\phi \cos \theta +\frac{k}{2})\Pi_{+} 
\end{eqnarray}
and
\be
\Pi_\pm := \frac{1}{2}(1\mp i e^{i \th \g_5} \g^i n_i).
\ee

\subsection{Fermi Condensate from Weyl Anomaly}

In this subsection, we use heat-kernel method
\cite{Vassilevich:2003xt} to derive Weyl anomaly due to a background
scalar.  To apply the heat-kernel method, we need to construct a
Laplace-type operator from the Dirac operator.
Following \cite{Parker:2009uva}, we define two operators
\begin{eqnarray}\label{D}
&&D=i \gamma^{i} \nabla_{i} +\phi ,\\ &&\widetilde{D}=-i \gamma^{i}
  \nabla_{i} +\phi. \label{bD}
\end{eqnarray}
In even dimensions, $\{\gamma^i\}$ and $\{-\gamma^i\}$ form equivalent
representations of Clifford algebra \cite{Parker:2009uva}. As a
result, the effective action can be rewritten as
\begin{eqnarray}\label{W}
W=-i \ln \text{det} D=-\frac{i}{2} \ln \text{det} (\widetilde{D}D),
\end{eqnarray}
where
\begin{eqnarray}\label{DD}
  \widetilde{D}D
  =g^{ij}\nabla_{i}\nabla_{j}+\frac{1}{4}R+\phi^2-i
  \gamma^{i} \partial_{i} \phi
  := g^{ij}\nabla_{i}\nabla_{j}-E,
\end{eqnarray}
where
\begin{eqnarray}\label{E}
E :=-\frac{1}{4}R-\phi^2+i \gamma^{i} \partial_{i} \phi.
\end{eqnarray}

Now we are ready to derive Weyl anomaly. Using the heat kernel
coefficient in \cite{Vassilevich:2003xt},
the Weyl anomaly related to the background scalar is
given by
\begin{eqnarray}\label{trE2}
\cA&=&\frac{1}{360(4\pi)^2} \int_M dx^4\sqrt{|g|}
\Big(-60\Box E+60 R E+180 E^2 \Big)\nonumber\\ &+&\frac{1}{360(4\pi)^2}
\int_M dy^3\sqrt{|h|} \Big( -(240 \Pi_{+}-120\Pi_-) n^i
\nabla_i E + 120 E k
+720 S E +120 S R\nonumber\\ &&\ \ \ \ \ \ \ \ \ \ \ \ \ \ \ \ \ +144
S k^2 +48 S k_{ab}k^{ab}+480 S^2 k+480 S^3+120 S D_a\chi D^a\chi \Big)
\end{eqnarray}
where $D_a$ denote covariant derivative on the boundary and we have
change the sign of $\Box E$, $n^i
\nabla_i E $ and $S D_a\chi D^a\chi$ of \cite{Vassilevich:2003xt} due to
different choice of signature in this paper. 
Substituting \eq{chi}, \eq{S}, \eq{E} and 
$D_a\chi=-i e^{i \theta
  \gamma_5} \gamma^i k_{ai}$ into (\ref{trE2}), we
obtain
\begin{eqnarray}\label{A}
\mathcal{A}=\frac{1}{8\pi^2}\Big( \int_M\sqrt{|g|} \left( -(\nabla
\phi)^2 +\frac{1}{6} R\phi^2+ \phi^4\right) +\int_{\partial
  M}\sqrt{|h|} \frac{1}{3} k\phi^2 \Big)+\sum_{n=1}^{4} b_n
\mathcal{B}_n
\end{eqnarray}
where $\mathcal{B}_m$ are given by
(\ref{B1},\ref{B2},\ref{B3},\ref{B4}) and $b_m$ are boundary central
charges,
 \begin{eqnarray}\label{centralcharges}
   b_1= \frac{\cos\theta-\frac{2}{3}\cos^3\theta}{4\pi^2},\quad
   b_2= -\frac{\cos^2\theta}{12\pi^2}, \quad
   b_3= \frac{\cos\theta}{48\pi^2},\quad
   b_4= \frac{\cos\theta}{40\pi^2}.
 \end{eqnarray}
It is remarkable that the Weyl anomaly (\ref{A}) for general BC
(\ref{chiralBC}) is Weyl invariant. This can be regarded as a check of
our calculations. Besides, for $\theta=\pm \frac{\pi}{2}$, all the
boundary central charges vanish and
\eq{A} reduces to the Weyl anomaly of \cite{Chu:2020mwx}.
For general BCs, the boundary central
charges (\ref{centralcharges}) are no longer zero.  This leads to
Fermi condensation $\la\bar{\psi}\psi\ra \sim \frac{1}{x^3}+ \cdots$ from
(\ref{FermicondensationGoodI}). In  the case of flat space with a flat
boundary, i.e. $R_{ijkl}=k_{ij}=0$, the Fermi
condensation (\ref{FermicondensationGoodI}) (\ref{centralcharges}) can
be simplified as
\begin{eqnarray}\label{condensation}
  \la\bar{\psi}\psi\ra
  = \frac{\cos\theta}{4\pi^2}\frac{1}{x^3}
  -\frac{\cos^2\theta}{4\pi^2}\frac{\phi(x)}{x^2}
  +\frac{1}{4\pi^2}\frac{
    \partial_x \phi(x) -(3\cos\theta-2\cos^3\theta)\phi^2(x) }{x}+ \cdots
  \, .
\end{eqnarray}

\subsection{Fermi Condensate from Green Function Method}

In this subsection, we study the anomalous Fermi condensation near a
boundary by applying the Green's function method \cite{Hu:2020puq}. For
simplicity, we focus on the linear order of background scalar. We
verify the result (\ref{condensation}) in a flat half
space.

Following \cite{Hu:2020puq} , let us first derive the Green's function at
the linear order of the background scalar field.  Green's function of the
Dirac fields satisfies
\begin{eqnarray}\label{GreenEOMDirac}
\left( i \gamma^{i} \nabla_i+\phi\right) S(x,x')=\delta(x,x'),
\end{eqnarray}
where $\delta(x,x'):=\delta^4(x-x')/\sqrt{|g|}$.  We impose the BCs
(\ref{generalBBC})
\begin{eqnarray}\label{GreenBBC}
\Pi_{-} S(x,x')|_{\partial M}=0,
\end{eqnarray}
where $\chi$ is given by \eq{chi}. $S$ also satisfies
\begin{eqnarray}\label{GreenBBC1}
S(x'',x) \gamma^n S(x,x')|_{\partial M}=0,
\end{eqnarray}
which follows immediately from \eq{GreenBBC} and  \eq{gamma}.
To solve for $S$ perturbatively, let us
split the Green's function into the background term $S_0$ and a correction
term $S_c$,
\begin{eqnarray}  \label{GreentotalDirac}
S=S_0+S_c,
\end{eqnarray}
where $S_0$ obeys the EOM 
\begin{eqnarray}\label{S0EOM}
i \gamma^{i} \nabla_{i} S_0(x,x')=\delta(x,x')
\end{eqnarray}
and the BC
\begin{eqnarray}\label{S0BBC}
\Pi_{-} S_0(x,x')|_{\partial M}=0.
\end{eqnarray}
For reasons similar to that of \eq{GreenBBC1}, it is easy to see that
\begin{eqnarray}\label{BBCBBC}
S_A(x'',x) \gamma^n S_B(x,x')|_{\partial M}=0.
\end{eqnarray}
where $S_{A,B}$ denotes $S,S_0, S_c$.
Let us apply the Green's formula for Dirac fields. We obtain
\begin{eqnarray}  \label{GreenformulaDirac}
&&\int_M d^4x\sqrt{|g|} \Big[ S_c(x', x) ( i \gamma^{i}
    \overrightarrow{\nabla}_{i}+\phi ) S(x, x'') + S_c(x', x) (i
    \gamma^{i} \overleftarrow{\nabla}_{i}-\phi ) S(x, x'')
    \Big]\nonumber\\ &=&
- \int_{\partial M} d^3x\sqrt{|h|} \Big[
    S_c(x', x) i \gamma^n S(x, x'') \Big] =0 ,
\end{eqnarray}
where $\overleftarrow{\nabla}_{i}$ means acting on the left and we
have used \eq{BBCBBC} in the last equation above.
Now \eq{GreenEOMDirac} and \eq{S0EOM} imply that
\begin{eqnarray}\label{ScEOM}
S_c(x', x) (i \gamma^{i} \overleftarrow{\nabla}_{i}-\phi )=S_0(x',x)
\phi(x).
\end{eqnarray}
Substituting to \eq{GreenformulaDirac}, we obtain the integral equation for $S_c$
\begin{eqnarray}\label{ScSolution}
S_c(x', x'')=-\int_M d^4x\sqrt{|g|} \Big[ S_0(x', x) \phi(x) S(x,x'')
  \Big],
\end{eqnarray} 
 and perturbatively we have
\begin{eqnarray}\label{Scper}
S_c(x', x'')&=&-\int_M d^4x\sqrt{|g|} S_0(x', x) \phi(x) S_0(x,x'')
\nonumber\\ &+& \int_M d^4x \sqrt{|g|} \int_{M_1} d^4x_1 \sqrt{|g_1|}
S_0(x', x)\phi(x) S_0(x,x_1)\phi(x_1) S_0(x_1,x'') \nonumber\\ &+& \cdots
\end{eqnarray} 
where the $n$-th line of (\ref{Scper}) is of order $O(\phi^n)$.

The Feynman Green function of Dirac field is given by
\cite{Parker:2009uva}
\begin{eqnarray}\label{GreenDirac}
S(x,x')=-i \la T\psi(x)\bar{\psi}(x')\ra,
\end{eqnarray}
where $T$ is the time-ordering symbol. From (\ref{GreenDirac}) one can
derive the Fermi condensation
\begin{eqnarray}\label{regCurrentDirac}
\la \bar{\psi} \psi \ra=-i\lim_{x'\to x}
\text{Tr}\Big[S(x,x')-\bar{S}(x,x')\Big],
\end{eqnarray}
where we have subtracted the reference Green function $\bar{S}$
for the theory without boundary.  From the key formula (\ref{Scper}), we get
\begin{eqnarray}\label{SDirac}
S(x', x'')=S_0(x', x'')- \int_0^{\infty}dx
\int_{-\infty}^{\infty}dtd^2y S_0(x', x) \phi(x)
S_0(x,x'')+O(\phi^2), \\ \bar{S}(x', x'')=\bar{S}_0(x', x'')-
\int_{-\infty}^{\infty}dx \int_{-\infty}^{\infty}dtd^2y
\bar{S}_0(x', x) \phi(x) \bar{S}_0(x,x'')+O(\phi^2). \label{barSDirac}
\end{eqnarray} 
where $\phi(x)=\phi_0+ x\phi_1$ and
\begin{eqnarray}\label{S0Dirac}
&&S_0(x', x'')=\frac{1}{2 \pi ^2}\Big[ \frac{\gamma^0(t'-t'')
    -\gamma^1 (x'-x'')-\gamma^a
    (y'_a-y''_a)}{((x'-x'')^2+(y'_a-y''_a)^2-(t'-t'')^2)^{2}}\nonumber\\
  &&\ \ \ \ \ \ \ \ \ \ \ \ \ \ \ \ \ \ \ \ \ \ \ \ \ \ \ \ \ \ \ \ \
  +\chi. \frac{\gamma^0(t'-t'')
    -\gamma^1 (-x'-x'')-\gamma^a
    (y'_a-y''_a)}{((x'+x'')^2+(y'_a-y''_a)^2-(t'-t'')^2)^{2}}\Big],
  \\
  && \bar{S}_0(x', x'')=\frac{1}{2 \pi ^2}
  \frac{\gamma^0(t'-t'') -\gamma^1 (x'-x'')-\gamma^a
    (y'_a-y''_a)}{[(x'-x'')^2+(y'_a-y''_a)^2-(t'-t'')^2]^{2}}.\label{barS0Dirac}
\end{eqnarray} 
Note that the integration region of $x$ are different for $S$ and
$\bar{S}$.  Substituting \eq{chi}, \eq{SDirac}-\eq{barS0Dirac}
into (\ref{regCurrentDirac}) and performing the Wick rotation $t=-i t_E$,
we obtain
\begin{eqnarray}\label{Diracintegral1}
\la\bar{\psi}\psi\ra= \frac{\cos \theta}{4 \pi ^2
  x^3}-\int_{0}^{\infty} dx' \int_0^{\infty} dr \frac{4 r^2 \left(\cos
  (2 \theta ) \left(\phi _1 x'+\phi _0\right)-\phi _1 x'+\phi
  _0\right)}{\pi ^3 \left(r^2+\left(x'+x\right)^2\right)^3}+O(\phi^2),
\end{eqnarray} 
where we have performed the angular integrals above. Carrying out the integrals
along $x'$ and $r$, we obtain the anomalous Fermi condensation in a
half space \begin{eqnarray}\label{Diraccondensation}
  \la\bar{\psi}\psi\ra&=& \frac{\cos \theta}{4 \pi ^2 x^3}-\frac{
    \cos ^2\theta \phi(x)}{4 \pi ^2 x^2}+\frac{\del_x \phi(x)}{4 \pi ^2
    x}+O(\phi^2),
\end{eqnarray} 
which agree with (\ref{condensation}) precisely. 

Following the same approach, we can derive the axial vector current
\begin{eqnarray}\label{chiralcurrent}
\la\bar{\psi}\gamma^5 \gamma^i\psi\ra&=& \Big[\frac{\sin \theta \phi(x)}{4 \pi ^2
  x^2}+\frac{\sin \theta \del_x \phi(x)}{4 \pi ^2
  x}+O(\phi^2) \Big] \d^i_1
\end{eqnarray} 
and the pseudo-Fermi condensation
\begin{eqnarray}\label{pseudo-fermi}
\la\bar{\psi} i \gamma^5\psi\ra&=&-\frac{\sin\theta}{4 \pi ^2
  x^3}
+\frac{\sin (2 \theta ) \phi (x) }{8 \pi ^2  x^2}
+ O(\phi^2).
\end{eqnarray} 
It is interesting that the normal axial vector current and
pseudo-Fermi condensation 
are non-zero for chiral angle $\theta\ne 0$.

\subsection{Condensation due to Pseudoscalar}

In this subsection, we generalize the above discussions to include 
Yukawa coupling with pseudoscalar.
Since the calculations are similar to those of section
4.1 and section 4.2, we will list only the key steps and key results below.

Let us start with the action
\begin{eqnarray} \label{Fermiactionpseudo}
I=\int_M \sqrt{|g|} \bar{\psi}\left(i \gamma^{i}\nabla_{i}+ \phi + i
\gamma_5 \bar{\phi} \right) \psi,
\end{eqnarray}
where $ \phi$ and $\bar{ \phi}$ are background scalar and
pseudoscalar, respectively.  Following section 4.1, we construct two
operators
\begin{eqnarray}\label{Dpseudo}
&&D=i \gamma^{i} \nabla_{i} +\phi+ i \gamma_5
  \bar{\phi},\\ &&\widetilde{D}=-i \gamma^{i} \nabla_{i} +\phi -i
  \gamma_5 \bar{\phi}. \label{bDpseudo}
\end{eqnarray}
Since $\{\gamma^i, \gamma_5\}$ and $\{-\gamma^i, -\gamma_5\}$ form
equivalent representations of Clifford algebra in even dimensions
\cite{Parker:2009uva}, we have
\begin{eqnarray}\label{Wpseudo}
W=-i \ln \text{det} D=-\frac{i}{2} \ln \text{det} (\widetilde{D}D).
\end{eqnarray}
From \eq{Dpseudo}, \eq{bDpseudo} and
$\widetilde{D}D= g^{ij}\nabla_{i}\nabla_{j}-E$, we get
\begin{eqnarray}\label{Epseudo}
E = -\frac{1}{4}R-\phi^2+i \gamma^{i} \partial_{i}
\phi-\bar{\phi}^2+\gamma_5 \gamma^{i} \partial_{i} \bar{\phi}.
\end{eqnarray}
Following the approach of section 4.1, we obtain
\begin{eqnarray}\label{formulapseudo}
S=-(\frac{k}{2}+\phi \cos\theta-\bar{\phi} \sin\theta)
\Pi_+,\ \ \ \chi=-i e^{i \theta \gamma_5} \gamma^i n_i,\ \ \ D_a \chi=-i e^{i
  \theta \gamma_5} k_{ab}\gamma^b.
\end{eqnarray}
Substituting \eq{Epseudo} and \eq{formulapseudo} into (\ref{trE2}),
we obtain the Weyl anomaly
\begin{eqnarray}\label{Apseudo}
  \mathcal{A}&=& a_1\mathcal{A}_1(\phi)
  +\bar{a_1}\mathcal{A}_1(\bar{\phi})+\sum_{n=1}^{4}
b_n \mathcal{B}_n(\phi)+\sum_{n=1}^{4} \bar{b_n}
\mathcal{B}_n(\bar{\phi})\nonumber\\ &&+\frac{1}{8\pi^2}\int_M
\sqrt{|g|} (\phi^2+\bar{\phi}^2)^2 +b_0
\int_{\partial M} \sqrt{|h|} \left( k \phi\bar{\phi}+\frac{3}{2}
n^i\nabla_i (\phi\bar{\phi})\right)
\end{eqnarray}
where $\mathcal{A}_1, \mathcal{B}_m$ are defined by
\eq{A1}, \eq{B1}-\eq{B4}; 
$a_1=\bar{a}_1=\frac{1}{8\pi^2}$, $b_m$'s are given by
(\ref{centralcharges}), $\bar{b}_m$'s are boundary central charges
related to the pseudoscalar
 \begin{eqnarray}\label{centralchargespseudo}
\bar{ b}_1 =
-\frac{\sin\theta-\frac{2}{3}\sin^3\theta}{4\pi^2},\quad
\bar{ b}_2 =-\frac{\sin^2\theta}{12\pi^2}, \quad
\bar{ b}_3 =  - \frac{\sin\theta}{48\pi^2},\quad
\bar{  b}_4 = -\frac{\sin\theta}{40\pi^2}
 \end{eqnarray}
 and
 \be
b_0 = \frac{\sin(2\theta)}{12\pi^2}
\ee
is the central charge associated with the last (new) anomaly term in \eq{Apseudo}. 
 It is interesting that the boundary central charge obeys the
 following relation
  \begin{eqnarray}\label{relationpseudo}
\bar{ b}_m (\theta)= b_m (\theta +\frac{\pi}{2}), \quad m = 1, 2, 3, 4.
 \end{eqnarray}
 Besides, (\ref{Apseudo}) is Weyl invariant, which can be regarded as
 a test of our calculations.

From the Weyl anomaly (\ref{Apseudo}) and the key formula
\begin{eqnarray} \label{key-1}
 (\delta_{\phi, \bar{\phi}} \mathcal{A})_{\partial M}=\left(\int_M
  \sqrt{|g|} \left(\la\bar{\psi}\psi\ra \delta \phi +
  \la\bar{\psi} i \gamma_5 \psi \ra \delta \bar{\phi}
  \right)\right)_{\log\epsilon},
\end{eqnarray}
one can derive the Fermi condensate
\be \label{FermicondensationpseudoI}
  \la\bar{\psi} \psi\ra = \mbox{RHS of  \eq{FermicondensationGoodI}} +
    \frac{\frac{3}{2}b_0 \bar{\phi}(x)}{x^2}+
    \frac{\frac{1}{2}b_0 k \bar{\phi}(x)}{x}
\ee
and the pseudo-Fermi condensation
\be\label{FermicondensationpseudoII}
\la\bar{\psi} i \gamma_5 \psi\ra = \mbox{RHS of \eq{FermicondensationpseudoI} with
  $(\phi, \bar{\phi}, b_m)$ replaced by $(\bar{\phi}, \phi, \bar{b}_m)$}.
\ee
It is
interesting that the pseudoscalar can induce Fermi condensation and
similarly the scalar can induce pseudo-Fermi condensation. In a flat
half space, the Fermi condensation (\ref{FermicondensationpseudoI})
and the pseudo-Fermi condensation (\ref{FermicondensationpseudoII})
becomes
\begin{eqnarray}\label{flatcondensationpseudoI}
  \la\bar{\psi}\psi\ra=\frac{\cos\theta}{4\pi^2}\frac{1}{x^3}
  -\frac{\cos^2\theta}{4\pi^2}\frac{
  \phi(x)}{x^2}+\frac{\sin(2\theta)}{8\pi^2}\frac{
  \bar{\phi}(x)}{x^2}+\frac{\partial_x
  \phi(x)-\phi^2(x) (3\cos\theta-2\cos^3\theta)}{4\pi^2x}+ \cdots, 
\end{eqnarray}
\begin{eqnarray}\label{flatcondensationpseudoII}
\la\bar{\psi}i
\gamma_5\psi\ra=\frac{-\sin\theta}{4\pi^2}\frac{1}{x^3}-
\frac{\sin^2\theta}{4\pi^2}\frac{
  \bar{\phi}(x)}{x^2}+\frac{\sin(2\theta)}{8\pi^2}\frac{
  \phi(x)}{x^2}+\frac{\partial_x
  \bar{\phi}(x)+\bar{\phi}^2(x)(3\sin\theta-2\sin^3\theta)}{4\pi^2x}+ \cdots .
\end{eqnarray}
Similar to section 4.2, one can verify (\ref{flatcondensationpseudoI})
and (\ref{flatcondensationpseudoII}) by applying Green's function
method.  The methods are the same as those of section 4.2, except that
one needs to replace $S_c$ by the following one
\begin{eqnarray}\label{Scperpseudo}
S_c(x', x'')=-\int_M d^4x \sqrt{|g|} S_0(x', x)\left(\phi(x) + i
\gamma_5 \bar{\phi}(x)\right) S_0(x,x'') + O(\phi^2,
\bar{\phi}^2,\phi\bar{\phi}).
\end{eqnarray}

\section{Holographic Story I: CFT with Boundary}

In this section, we study the one point function of scalar operator
$O$ in holographic BCFT \cite{Takayanagi:2011zk}.  
We will derive the
holographic one point functions
and holographic Weyl anomaly
and find that they indeed
obey the universal relations (\ref{results1}) between Fermi
condensation and central charges.
For our purpose, it will be sufficient to consider the Euclidean version
of the AdS/CFT correspondence. Anomalies and correlation functions
in zero temperature Minkowski theory can be obtained directly
by Wick rotation. 
Note that, we use signature $(1,1,1,1)$ instead of (1,-1,-1,-1) in
this and the next section. It should be mentioned that the one point
function 
(e.g. Fermi condensation) is independent of the choice of  signature.

\begin{figure}[t]
\centering \includegraphics[width=5cm]{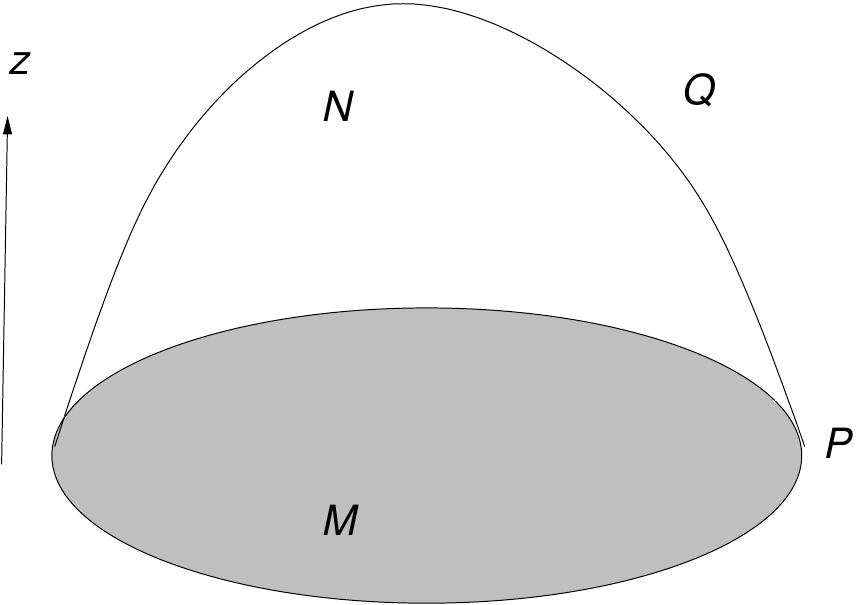}
\caption{BCFT on $M$ and its dual $N$}
\label{MNPQ}
\end{figure}

Let us first give a quick review of the geometry of holographic BCFT
\cite{Takayanagi:2011zk}.  Consider a BCFT \cite{Cardy:2004hm} defined
on a manifold $M$ with a boundary $P$.  Takayanagi
\cite{Takayanagi:2011zk} proposed to extend the $d$-dimensional
manifold $M$ to a $(d+1)$-dimensional asymptotically AdS space $N$
such that $\partial N= M\cup Q$, where $Q$ is a $d$ dimensional
manifold with boundary $\partial Q=\partial M=P$. See figure
\ref{MNPQ} for example.

Without loss of generality, we choose the following bulk action in this paper
\begin{eqnarray} \label{action}
I=\int_N d^5x\sqrt{|G|}
\left(\hat{R}+12-\frac{1}{2}\left(\hat{\nabla}^{\mu}
\hat{\phi}\hat{\nabla}_{\mu} \hat{\phi}+ m^2 \hat{\phi}^2\right)
\right)+2\int_Q dx^4\sqrt{|\gamma|} \left(K-T +\frac{\xi}{2}
\hat{\phi} \right),
\end{eqnarray}
where we have set $16 \pi G_N=1$ and AdS radius $l=1$ for simplicity. 
Note that the Euclidean action is given by $I_E=-I$ with signature $(1,1,1,1)$.
Here ($G_{\mu\nu}$, $\hat{R}$, $\hat{\nabla}_{\mu}$, $\hat{\phi}$) are the
metric, scalar, covariant derivatives and Ricci scalar in the bulk
$N$, ($\gamma_{ij}, K$) are the induced metric and extrinsic curvature
on the bulk boundary $Q$, $m$ is the mass of scalar field $\hat{\phi}$ and ($T,
\xi $) are constant parameters of the theory. Note that $T$  can be
regarded as holographic dual of the boundary entropy
\cite{Takayanagi:2011zk,Miao:2017gyt,Chu:2017aab}, while,
as we will see later that, $\xi$
parameterizes the boundary condition of the scalar field.
To have a well-defined action
principle, one must impose suitable boundary conditions on
$Q$. Following \cite{Takayanagi:2011zk}, we choose Neumann boundary
conditions (NBC)
\begin{eqnarray} \label{NBCmetric}
&&K_{ij}-(K-T+\frac{\xi}{2} \hat{\phi})\gamma_{ij} =0,\\ &&
  \hat{n}^{\mu}\hat{\nabla}_{\mu} \hat{\phi}-
  \xi=0, \label{NBCscalar}
\end{eqnarray}
where $\hat{n}^{\mu}$ is the outward-pointing normal vector on $Q$. Note
that there are other choices of consistent boundary conditions
\cite{Miao:2017gyt,Chu:2017aab,Miao:2018qkc}, which we leave for
future studies.  From the action (\ref{action}), we get equations of
motion (EOM)
\begin{eqnarray} \label{EOMmetric}
&& \hat{R}_{\mu\nu}-\frac{\hat{R}+12}{2} G_{\mu\nu}=\frac{1}{2}
  T_{\mu\nu},\\ &&
  (\hat{\nabla}^{\mu}\hat{\nabla}_{\mu}-m^2)\hat{\phi}=0, \label{EOMscalar}
\end{eqnarray}
where $T_{\mu\nu}$ is the stress tensor of the scalar field
\begin{eqnarray} \label{Tuvscalar}
T_{\mu\nu}=\hat{\nabla}_{\mu} \hat{\phi}\hat{\nabla}_{\nu}
\hat{\phi}-\frac{1}{2}G_{\mu\nu}\left(\hat{\nabla}^{\alpha}
\hat{\phi}\hat{\nabla}_{\alpha} \hat{\phi}+ m^2 \hat{\phi}^2\right).
\end{eqnarray}

Near the AdS boundary,
the scalar field behaves as
\begin{eqnarray} \label{AdSbdyscalar}
 \hat{\phi}=z^{4-\Delta}\phi(x) + +z^{\Delta} \phi_{(2\Delta -4)}(x), \quad z\to 0,
\end{eqnarray}
where $\phi$ is the boundary scalar discussed in section 2 and section
3, $\Delta=2+\sqrt{4+m^2}$ is the conformal dimension of the operator
$O$ dual to $\hat{\phi}$. According to the dictionary of AdS/CFT
\cite{Klebanov:1999tb,deHaro:2000vlm}, we have
\begin{eqnarray} \label{scalarcurrent}
  \la O \ra=\frac{1}{\sqrt{|g|}}\frac{\delta I}{\delta \phi}
  =(2\Delta-4)\phi_{(2\Delta -4)}+ \cdots
\end{eqnarray}
where $ \cdots$ denote finite and local functions of $(\phi, g_{ij}, \psi_{(2\Delta
  -4)})$. Since we are interested in the `divergent
terms' (\ref{FermicondensationI0}) near the boundary, we can ignore
these irrelevant $\cdots$ terms. 
For our purpose, we focus on the case
$\Delta =3$, or equivalently,
\begin{eqnarray} \label{MMM}
m^2=-3,
\end{eqnarray}
which is above the Breitenlohner-Freedman
stability bound $m^2 > -4$ for asymptotic AdS${}_5$. 

Now the approach to derive the holographic one point function is
straightforward. First we solve the coupled Einstein-scalar EOM
\eq{EOMmetric} and \eq{EOMscalar} with the boundary conditions
\eq{NBCmetric} and \eq{NBCscalar}. Then we use the scalar
solution to obtain the
holographic one point function \eq{scalarcurrent} from the
asymptotic behaviour \eq{AdSbdyscalar}.

It is a non-trivial problem to find solutions which satisfy the EOM
with the specified form of boundary conditions (BC).
For examples, the usual AdS black holes
are no longer solutions to AdS/BCFT generally, since they do not obey
NBC (\ref{NBCmetric}). A systematic method based on derivative expansion was
developed in \cite{Miao:2017aba,Chu:2018ntx,Miao:2018qkc}. Following
\cite{Miao:2017aba,Chu:2018ntx,Miao:2018qkc}, we take the following
ansatz for the bulk metrics
\begin{eqnarray}\label{bulkmetric}
 ds^2&=&\frac{1}{z^2}\Big{[} dz^2+ \left(1+\epsilon x
   X_1(\frac{z}{x})+\epsilon^2 x^2 X_2(\frac{z}{x})+\cdots\right)dx^2
   \nonumber \\ && +\left(\delta_{ab}-2\epsilon x
   K_{ab}(\frac{z}{x})+\epsilon^2 x^2 Q_{ab}(\frac{z}{x})
   +\cdots\right)dy^a dy^b\Big{]} \nonumber\\ &&+ O(\xi^2)
\end{eqnarray}
and the bulk scalar field
\begin{eqnarray}\label{bulkscalar}
\hat{\phi}=f_0(\frac{z}{x})+ \epsilon\ x f_1(\frac{z}{x})+ \epsilon^2
x^2 f_2(\frac{z}{x})+ \epsilon^3 x^3 f_3(\frac{z}{x})+ \cdots, 
\end{eqnarray}
where $X_n, K_{ab}, Q_{ab}, f_n$ are unknown functions to be
determined and $\xi$ is the parameter for the scalar boundary condition
\eq{NBCscalar}.
Note that we have introduced a parameter $\epsilon$ to label the
order of derivative expansions with respect to  $x$ or $z$.
It should be set $\e =1$ at the end of
calculations. To get an asymptotic AdS background, we set the BC
\begin{eqnarray}\label{fXQ}
& X_1(0)=X_2(0)=0, \quad K_{ab}(0)=k_{ab}, \quad Q_{ab}(0)=q_{ab},
  \nonumber\\
  & \lim_{z\to 0} \frac{f_0 (z)}{z}=0, \
  \lim_{z\to 0} \frac{f_1 (z)}{z}=\phi_0, \quad
  \lim_{z\to 0} \frac{f_2 (z)}{z}=\phi_1, \quad
  \lim_{z\to 0} \frac{f_3 (z)}{z}=\frac{1}{2}\phi_2, \quad \mbox{etc.},
\end{eqnarray}
so that the metric and scalar on $M$ take expected forms in the Gauss
normal coordinates
\begin{eqnarray} \label{metricGNC}
&&ds_M^2=dx^2+ \left(\delta_{ab}-2\epsilon \ x k_{ab}+\epsilon^2
  x^2q_{ab}+ \cdots \right) dy^ady^b,\\
  && \phi= \epsilon \phi_0+ \epsilon^2 x \phi_1+\epsilon^3 \frac{x^2}{2}\phi_2
  + \cdots .  \label{scalarGNC}
\end{eqnarray}
The powers of $\e$ in \eq{scalarGNC} is understood from the fact that $\phi$, being
the coefficient of $\hat{\phi}$ near $z=0$ as dedicated by \eq{AdSbdyscalar},
is already of order $\e$. 
We also take the embedding function of bulk boundary $Q$ to be of the form
\begin{eqnarray}\label{bulkQ}
x=-\sinh \rho z +\epsilon \lambda_1 z^2 +\epsilon^2 \lambda_2 z^3+ \cdots
\end{eqnarray}
where $\lambda_n$ are constants.  Note that functions $X_m, K_{ab},
Q_{ab}, f_n, \lambda_n$ are functions of $\xi$.
 
\subsection{Holographic Condensate}

Let us first study the background solution with
$\epsilon=0$. Substituting (\ref{bulkscalar}) into EOM
(\ref{EOMscalar},\ref{MMM}) , we get
\begin{eqnarray}\label{f0EOM}
\left(s^2+1\right) s^2 f_0''(s)+\left(2 s^2-3\right) s f_0'(s)+3
f_0(s)=0,
\end{eqnarray}
which has the solution
\begin{eqnarray}\label{f0Sol}
 f_0(s)=d_1 \xi \frac{ s^3}{\left(s^2+1\right)^{3/2}} +d_2 \xi
  \frac{s \left(\sqrt{s^2+1}+s^2 \tanh
   ^{-1}(\sqrt{s^2+1})\right)}{\left(s^2+1\right)^{3/2}},
\end{eqnarray}
where $d_1,d_2$ are integral constants and $s=\frac{z}{x}$. Imposing
the NBC (\ref{NBCscalar}) on the bulk boundary $Q$ and DBC (\ref{fXQ}) on AdS
boundary $M$, we fix the integral constants to be
\begin{eqnarray}\label{f0Sol1}
d_1=-\frac{1}{3}\cosh ^3\rho \, \coth\rho , \ \ \ d_2=0.
\end{eqnarray}
Thus, the scalar $f_0$ is of order $O(\xi)$. As a result, the
scalar stress tensor (\ref{Tuvscalar}) and thus the back reaction to
the bulk geometry is of order $O(\xi^2)$. This means that AdS
metric is a solution to (\ref{EOMmetric}) only up to
$O(\xi^2)$. That is the reason why we add $O(\xi^2)$ in
the last line of bulk metric (\ref{bulkmetric}). For simplicity, we
mainly focus on solutions up to $O(\xi)$ in this paper. We
discuss briefly
the effects of backreaction up to order $ O(\xi^2)$ to the metric and
$O(\xi^3)$ to the scalar field $\hat{\phi}$ in the appendix B.

Now we are ready to derive the leading term of one point function.
From \eq{scalarcurrent}, \eq{bulkscalar} and \eq{f0Sol}, we obtain
\begin{eqnarray} \label{O0}
\la O \ra =\frac{2 d_1 \xi}{x^3}+O(1/x^2, \xi^2).
\end{eqnarray}
Comparing with (\ref{FermicondensationGoodI}), we read off the central
charge
\begin{eqnarray} \label{b3}
b_3=\frac{d_1\xi}{6}= -\frac{\xi}{18} \cosh ^3\rho\, 
\coth \rho +O(\xi^2)
\end{eqnarray}

Following the same procedure, we can solve for the  bulk solutions
to \eq{bulkmetric} and \eq{bulkscalar} order in order in $\epsilon$ and
derive the sub-leading terms of the one point function. Since the
calculations are quite complicated, we will  first study below some special
cases and then list the general results. 
In following subsections, 
we will determine the bulk solution up to order $\e^2$ and linear order in $\xi$.

\subsubsection{Free-Field Limit}

To warm up, let us first study so-called Free-Field Limit. It is
noticed that, when the brane tension vanishes $T=0$, holographic Weyl
anomaly \cite{Astaneh:2017ghi}, norm of displacement operator
\cite{Miao:2018qkc,Miao:2018dvm} and their two point functions
\cite{Miao:2018dvm,Alishahiha:2011rg} all exactly match those of free
theories. So we call $T=0$ the free-field limit. When there are
scalars, a natural choice of the free-field limit would be to take
$\xi=0$ in addition to $T=0$. Equivalently, the boundary
conditions become
\begin{eqnarray} \label{freeNBCmetric}
&&K_{ij}-K\gamma_{ij} =0,\\ && \hat{n}^{\mu}\hat{\nabla}_{\mu}
  \hat{\phi}=0 \label{freeNBCscalar}.
\end{eqnarray}
Below we will show that the above boundary conditions can indeed
produce the form of one point function for free BCFT.

First from (\ref{bulkscalar},\ref{f0Sol},\ref{f0Sol1}), we notice that
$\hat{\phi}\sim O(\epsilon)$ when $\xi=0$. As a result, the
back reaction due to scalars to the bulk metric is of order
$O(\epsilon^2)$. Fortunately, to derive one point function up to
$O(\epsilon^2)$ ($O(1/x)$), we do not need the bulk metric of order
$O(\epsilon^2)$.  That is because, from EOM (\ref{EOMscalar}) and
$\hat{\phi}\sim O(\epsilon)$, the order $O(\epsilon^2)$ terms of the bulk metric
affect only the order $O(\e^3)$ terms of the bulk scalar and thus
are irrelevant for the one point function up to
order $O(\epsilon^2)$. This means we can ignore the back reaction of
scalars on the metric in the free-field limit $\xi=0$.

On this, we recall the metric without scalars were obtained in
\cite{Miao:2017aba},
where the bulk metric is given by
\begin{eqnarray}\label{bulkmetriclimit}
&& ds^2=\frac{1}{z^2}\Big{[} dz^2+ \big(1+x^2\epsilon^2
    X(\frac{z}{x})\big)dx^2 \nonumber \\ &&
    +\big(\delta_{ab}-2x\epsilon \bar{k}_{ab}
    f(\frac{z}{x})-2x\epsilon \frac{k}{3} \delta_{ab} +
    x^2\epsilon^2 Q_{ab}(\frac{z}{x}) \big)dy^a dy^b\Big{]}
  \nonumber\\ && +O(\epsilon^3),
\end{eqnarray}
and the embedding function of $Q$ is given by
\begin{eqnarray}\label{4dQlimit}
x=-\sinh(\rho) z+\epsilon \frac{k \cosh ^2\rho }{6} z^2 +\epsilon^2
\lambda_2 z^3+O(\epsilon^3)
\end{eqnarray}
Here $k_{ab}=\text{diag}(k_1,k_2,k_3)$, $T=3 \tanh\rho$, $f(s)$ is given by 
\begin{eqnarray}\label{flimit}
&&f(s)=1+2 \alpha _1-\frac{\alpha _1
    \left(s^2+2\right)}{\sqrt{s^2+1}},\\ &&\alpha_1=\frac{-1}{2 (1+\tanh
    \rho)},\nonumber
\end{eqnarray}
 and $X, Q_{ab},\lambda_2$ are complicated functions, which can be
 found in the appendix of \cite{Miao:2017aba} . As mentioned above, in
 the free-field limit, we do not need either $X, Q_{ab},\lambda_2$ which are of
 order $O(\epsilon^2)$ or a non-vanishing tension $T=3 \tanh\rho$. However,
 for the convenience of following sections, we will give the general
 results below by first studying the general case with $T=3 \tanh\rho$
 and then we will take the free-field limit $T\to 0$ at the end of
 calculations.
 
 Substituting bulk metric (\ref{bulkmetriclimit}) and scalar
 (\ref{bulkscalar}) with $f_0=0$ into EOM (\ref{EOMscalar}), we obtain
\begin{eqnarray}\label{f1free}
&&f_1(s)=s
  \left(\frac{d_3}{\sqrt{s^2+1}}+d_4\right),\\ &&f_2(s)=\frac{s
    \left(d_3 k+2 d_6 (s^2+1)\right)}{2 \sqrt{s^2+1}}+d_5
  s \label{f2free}.
\end{eqnarray}
Imposing DBC (\ref{fXQ}) on AdS boundary $z=0$, we get
\begin{eqnarray}\label{d4d6free}
d_4=\phi_0-d_3, \ \ d_6=-\frac{1}{2} d_3 k-d_5+\phi_1.
\end{eqnarray}
Imposing NBC (\ref{freeNBCscalar}) on bulk boundary $Q$, we obtain
\begin{eqnarray}\label{d3d5free}
  d_3=\frac{\phi _0 \coth \rho }{\coth \rho -\text{csch}^2\rho+1}, \quad
  d_5=\frac{k \phi _0 (-\sinh (2 \rho )+\cosh (2 \rho )-3)+3
  \phi _1 \sinh (2 \rho )}{3 (\sinh (2 \rho )+\cosh (2 \rho )-3)}.
\end{eqnarray}
Substituting bulk scalar solution (\ref{f1free}) and (\ref{f2free}) into
(\ref{scalarcurrent}) and (\ref{bulkscalar}), we obtain the one point
function
\begin{eqnarray}\label{Ofree0}
\la O \ra=-\frac{d_3}{x^2} \epsilon -\frac{ d_3 k+d_5-\phi _1}{x}\epsilon ^2
+O\left(\epsilon^3, \xi\right).
\end{eqnarray}
In the free-field limit $T=\rho=0$, it becomes
\begin{eqnarray}\label{Ofree}
\la O\ra =-\frac{ \left(\frac{k}{3} \phi _0- \phi _1\right)}{x} \epsilon
^2+O\left(\epsilon^3, \xi\right),
\end{eqnarray}
which takes the same form \eq{FermicondensationGoodI2}
as that of the free theories \cite{Chu:2020mwx}
with all the boundary charges vanish: $b_i =0$.
Comparing with (\ref{FermicondensationGoodI2}) and using $n^i\nabla_i
\phi=-\phi_1+O(x)$, we get the bulk central charge
\begin{eqnarray}\label{a1}
a_1=\frac{1}{2}.
\end{eqnarray}
Note that the bulk central charge is independent of boundary
conditions, so (\ref{a1}) is exact and gets no corrections from
$\epsilon$ and $\xi$.

\subsubsection{No-Scalar Limit}

Let us go on to investigate the no-scalar limit. By `no scalar' we
mean there is no boundary scalar, i.e., $\phi=0$, but the bulk scalar
$\hat{\phi}$ can be non-zero.  Now we have $\hat{\phi} \sim
O(\xi)$, which back react the bulk metric at order
$O(\xi^2)$. Since we mainly focus on solutions linear in
$\xi$, we can ignore the back reaction due to scalars to the
bulk metrics. Note that we have $\hat{\phi} \sim O(\xi)$ in
no-scalar limit, while $\hat{\phi} \sim O(\epsilon)$ in free-field
limit. As a result, unlike the case of free-field limit, in no-scalar
limit we need bulk metrics (\ref{bulkmetriclimit}) of order
$O(\epsilon^2)$ in order to get the one point function of order
$O(\epsilon^2)$.

Solving EOM (\ref{EOMscalar}) with bulk metric (\ref{bulkmetriclimit})
and impose the DBC (\ref{fXQ})  with $\phi_0=\phi_1=0$, we obtain $f_0$
(\ref{f0Sol}) with $d_2=0$ and
\begin{eqnarray}\label{f1noscalarlimit}
 f_1(s)=\frac{1}{2} d_1 k s
 \left(1-\frac{1}{\left(s^2+1\right)^{3/2}}\right)+e_1 s
 \left(\frac{1}{\sqrt{s^2+1}}-1\right)
 \end{eqnarray}
and
\begin{eqnarray}\label{f2noscalarlimit}
 f_2(s)&=&\frac{q \left(-\left(d_1 s^3 \left(3
   s^2+2\right)\right)\right)}{6 \left(s^2+1\right)^{5/2}}-\frac{k
   \left(e_1 s^3\right)}{2 \sqrt{s^2+1}}-\frac{e_2
   \left(s^3-\sqrt{s^2+1}
   s+s\right)}{\sqrt{s^2+1}}\nonumber\\ &&+\text{Tr}k^2
 \left(\frac{d_1 s^3 \left(5 s^4+21 s^2+14\right)}{8
   \left(s^2+1\right)^{5/2}}-3 h_1(s)\right)+k^2 h_1(s)
 \end{eqnarray}
 with
 \begin{eqnarray}\label{h1}
   h_1(s)&=&\frac{d_1 s}{360 \left( s^2+1\right)^{5/2}}
   \left[15 s^2(3 s^4+12 s^2+8)-30 \alpha_1^2 s^2 (3 s^2+2)
   \log \left(s^2+1\right)\right.\nonumber\\
   &&-12 \alpha_1 \left(2
   s^6-9 s^4+4 (5 \sqrt{s^2+1}-6) s^2+8
   (\sqrt{s^2+1}-1)\right)\nonumber\\ &&\left.+4\alpha_1^2
   \left(14 s^6+87 s^4-8 (15 \sqrt{s^2+1}-19) s^2-48
   (\sqrt{s^2+1}-1)\right) \right],
 \end{eqnarray}
 where $d_1$ is (\ref{f0Sol1}), $\alpha_1$ is given by (\ref{flimit}) and
 $e_1, e_2$ are integral constants.  Imposing the NBC (\ref{NBCscalar}),
 we fix the integral constants
\begin{eqnarray}\label{e1noscalarlimit}
e_1=-\xi \frac{k}{6} \cosh ^3\rho  \, \coth \rho 
 \end{eqnarray}
and
\begin{eqnarray}\label{e2noscalarlimit}
e_2&=&\xi\frac{\left(3 \text{Tr}k^2-k^2\right) \coth \rho \,
  {\rm csch}^2\rho }{1080 (\coth \rho -2) (\coth \rho +1)^2}
\nonumber\\
&&\left( 20 \sinh \rho +13 \sinh (3 \rho )+3 \sinh (5
\rho )-22 \cosh \rho -4 \cosh (3 \rho ) \right)
 \end{eqnarray}
 Substituting the above scalar solution into (\ref{scalarcurrent}), we
 obtain the one point function for $\phi=0$
 (\ref{FermicondensationGoodI3}) with the central charges $b_3$
 given by (\ref{b3}) and
 \begin{eqnarray} \label{b4}
b_4= - \xi\frac{-11 \sinh (2 \rho )-8 \sinh (4 \rho )+8 \cosh (2
  \rho )+7 \cosh (4 \rho )+1}{90 (-5 \sinh \rho +3 \sinh (3 \rho )+3
  \cosh \rho -3 \cosh (3 \rho ))}+O(\xi^2).
 \end{eqnarray}
 Note that there are four independent terms but only two parameters in
 (\ref{FermicondensationGoodI3}).  It is non-trivial to have
 consistent solutions (\ref{b3}) and (\ref{b4}). This is a strong support of
 our results.
 
 \subsubsection{Flat Limit}
 \label{flat-limit}
 
 In this subsection, we consider the back reaction of scalars. For
 simplicity, we focus on the flat space with flat boundary, i.e.,
 $k_{ab}=q_{ab}=0$. We denote this case as the `flat limit'.  Since
 the calculations are quite similar to those of above subsections,
 below we only show the key steps.
 
 In the flat limit, the ansatz for bulk metrics and bulk scalar can be
 simplified as
 \begin{eqnarray}\label{bulkmetricflatlimit}
 ds^2&=&\frac{1}{z^2}\Big{[} dz^2+ \left(1+\epsilon \xi x
   X_{1e}(\frac{z}{x})+\epsilon^2 x^2 \left(X_2(\frac{z}{x})+
   \xi X_{2e}(\frac{z}{x})\right)\right)dx^2 \nonumber \\ &&
   +\delta_{ab} \left(1+\epsilon \xi x
   g_{1e}(\frac{z}{x})+\epsilon^2 x^2 \left(
   g_{2}(\frac{z}{x})+\xi g_{2e}(\frac{z}{x})\right)
   \right)dy^a dy^b\Big{]} \nonumber\\ &&+ O(\xi^2, \epsilon^3)
\end{eqnarray}
and
\begin{eqnarray}\label{bulkscalarflatlimit}
\hat{\phi}=f_0(\frac{z}{x})+ \epsilon\ x f_1(\frac{z}{x}) + \epsilon^2
x^2 \left( f_2(\frac{z}{x}) +\xi f_{2e}(\frac{z}{x})
\right)+O(\xi^2, \epsilon^3),
\end{eqnarray}
where $f_0$ is given by (\ref{f0Sol},\ref{f0Sol1}) up to order
$O(\xi)$. Solving the coupled Einstein-scalar EOM
(\ref{EOMmetric}), (\ref{EOMscalar}) and the DBC (\ref{fXQ}) with
$k_{ab}=q_{ab}=0$, we obtain for the bulk scalar
\begin{eqnarray}\label{f1flatlimit}
&&f_1(s)=s \left(d_3 (\frac{1}{\sqrt{s^2+1}}-1)+\phi
  _0\right), \\ \label{f2flatlimit}
  &&f_2(s)=s \left(\sqrt{s^2+1}
  \left(\phi _1-d_5\right)+d_5\right)
\end{eqnarray}
and
\begin{eqnarray}\label{f2bflatlimit}
&&f_{2e}(s)= \frac{ s \left(-s^2+\sqrt{s^2+1}-1\right)}{\sqrt{s^2+1}}
  d_7\nonumber \\
  &&+\frac{d_1 s}{48 \left(s^2+1\right)^{5/2}} \Big[-2
    s^2 \left(15 (s^2+2) s^2+16\right) \phi _0^2 \nonumber
    \\ && +2 d_3 \left(25 s^6+33 s^4+(26 \sqrt{s^2+1}-15)
    s^2+17 (\sqrt{s^2+1}-1)\right) \phi _0 \nonumber \\ &&
    +d_3^2 \left(-19 s^6-52 \sqrt{s^2+1} s^2+54 s^2-34 \sqrt{s^2+1}-6
    s^4 \left(\log (s^2+1)+2\right)+34\right) \Big]. 
\end{eqnarray}
Imposing the NBC (\ref{NBCmetric}), (\ref{NBCscalar}), we obtain the integral
constants $d_1$ (\ref{f0Sol1}) and
\begin{eqnarray}\label{d3flatlimit}
&&d_3=\frac{-\phi _0 \sinh \rho  \cosh \rho }{(\cosh \rho -2
    \sinh \rho ) (\sinh \rho+\cosh \rho)},\\ \label{d5flatlimit}
  &&d_5=\frac{\phi _1 \sinh (2 \rho
    )}{\sinh (2 \rho )+\cosh (2 \rho )-3},\nonumber\\
  &&d_7=\frac{\phi_0^2\text{csch}^5\rho }{9216 (\coth \rho -2)^2 (\coth \rho
    +1)^2} \Big[688-1106 \sinh (2 \rho )-258 \sinh (4 \rho )+70 \sinh
    (6 \rho ) \nonumber\\ && \ \ \ \ +\sinh (8 \rho )+616 \cosh (2
    \rho )-192 \cosh (4 \rho )-104 \cosh (6 \rho )+16 \cosh (8 \rho )
    \Big].
\end{eqnarray}
Please see appendix B for the solutions to the bulk metric
(\ref{bulkmetricflatlimit}) and the embedding function of $Q$
(\ref{bulkQ}) . Substituting the above scalar solutions into
(\ref{scalarcurrent}), we obtain the one point function in flat limit
 \begin{eqnarray} \label{onepointflatlimit}
\la O \ra=\frac{12 b_3}{x^3}+\frac{3b_2 \phi}{x^2} \epsilon- \frac{
  2a_1\nabla_n \phi+3 b_1 \phi^2 }{x} \epsilon^2+O(x^0,\ln x),
 \end{eqnarray}
where $b_3$ is given by (\ref{b3}), $a_1$ is given by (\ref{a1}) and
  \begin{eqnarray} \label{b1}
&&b_1=-\xi \frac{ \text{csch}^3\rho }{1152 (\coth \rho
      -2)^2} \Big(44+57 \cosh (2 \rho )+20 \cosh (4 \rho
    )\nonumber\\ &&\ \ \ \ \ \ \ \ -61 \sinh (2 \rho )-30 \sinh (4
    \rho )-5 \sinh (6 \rho )+7 \cosh (6 \rho ) \Big) +O(\xi^2),
 \end{eqnarray}
 \begin{eqnarray} \label{b2}
   b_2= \frac{\sinh \rho  \cosh \rho }{3 (\cosh \rho -2 \sinh \rho
  ) (\sinh \rho +\cosh \rho)}+O(\xi^2).
 \end{eqnarray}
Note that $a_1$ and $b_3$ derived in the flat limit
(\ref{onepointflatlimit}) agree with those obtained in free-field
limit and no-scalar limit. This can be regarded as a double check of
our results.  Now we have got all of the boundary central charges
$b_1$ (\ref{b1}), $b_2$ (\ref{b2}), $b_3$ (\ref{b3}) and $b_4$
(\ref{b4}) in holographic BCFT (\ref{action}).

So far, we have verified the one point function
(\ref{FermicondensationGoodI}) in three special cases. The
generalization to general case is straightforward. However, the
general solutions to the bulk metric (\ref{bulkmetric}) and bulk
scalar (\ref{bulkscalar}) are quite complicated, we do not list them
in this paper. The interested reader can obtain them straightforwardly
with the help of Mathematica. Besides, we focus on solutions in the
linear order of $\xi$ in this section. Please refer to appendix
B for solutions in higher orders of $\xi$.

\subsection{Holographic Weyl Anomaly}

In this subsection, we investigate the holographic Weyl Anomaly due to
the scalar background. In particular, we reproduce the four boundary
central charges $b_1,b_2,b_3,b_4$ obtained in section 4.1 and verify
the universal relations between one point function
(\ref{FermicondensationGoodI}) and Weyl anomaly (\ref{anomaly}).

\subsubsection{Bulk Weyl Anomaly}

Let us first consider the bulk contributions to Weyl anomaly, where we
can ignore the boundaries.  For this case we can apply the standard
method \cite{Henningson:1998gx} to derive the holographic Weyl
Anomaly.  Due to the non-trivial back reactions, the case with
boundaries is more subtle, and we leave a careful study of it in next
subsection.

Following \cite{Henningson:1998gx}, we take the Fefferman-Graham gauge
for the asymptotically $AdS_5$ metric
\begin{eqnarray}\label{AdSmetric}
ds^2=\frac{d\rho^2}{4\rho^2}+\frac{\hat{g}_{ij}(x,\rho)dx^idx^j}{\rho},
\end{eqnarray}
where $\hat{g}_{ij}=g_{ij}+\rho g_{(1)ij}+ \cdots$ and $\rho=z^2$.  Using the
EOM (\ref{EOMmetric}) together with (\ref{AdSbdyscalar}) with
$\Delta=3$ and (\ref{AdSmetric}), we obtain
   \begin{eqnarray} \label{g1ij}
g^{(1)}_{ij}=-\frac{1}{2}R_{ij}+\frac{1}{12}(R-\phi^2) g_{ij}.
 \end{eqnarray}
Substituting the bulk metric (\ref{AdSmetric},\ref{g1ij}) and bulk
scalar (\ref{AdSbdyscalar}) with $\Delta=3$ into action
(\ref{action}), selecting UV logarithmic divergent terms, we obtain
the bulk contributions to holographic Weyl anomaly
\begin{eqnarray}\label{bulkanomalysignature2}
  \mathcal{\hat{A}}_{\text{bulk}}=( I)_{\log \frac{1}{\epsilon}}=
  \frac{-1}{2}\int_M \sqrt{g} [(\nabla
   \phi)^2+\frac{1}{6}R\phi^2+ \frac{1}{6}\phi^4]
 \end{eqnarray}
 in signature $(1,1,1,1)$ or $(-1,1,1,1)$. In signature
 $(1,-1,-1,-1)$, the definition of Weyl anomaly
 $\mathcal{A}=\la T^i_{\ i} \ra$ change sign. That is because $T^i_{\ i}$
 differs by a minus sign in different signature.  Please see appendix
 A for more clarifications.  Transform into signature $(1,-1,-1,-1)$,
 the bulk Weyl anomaly becomes
 \begin{eqnarray}\label{bulkanomaly}
 \mathcal{A}_{\text{bulk}}=( I)_{\log \epsilon}=\frac{1}{2}\int_M \sqrt{g} [-(\nabla
   \phi)^2+\frac{1}{6}R\phi^2+ \frac{1}{6}\phi^4],
 \end{eqnarray}
from which one can read off the bulk central charges
\begin{eqnarray}\label{bulkcentralcharge}
a_1=\frac{1}{2}, \quad a_2=\frac{1}{12}
 \end{eqnarray}
which agree with (\ref{a1}).  To avoid confusion, by central charges we
always
refer to those coefficients appearing in the Weyl anomaly (\ref{anomaly})
in signature $(1,-1,-1,-1)$.

\subsubsection{Boundary Weyl Anomaly}

Let us turn to discuss the boundary contributions to holographic Weyl
anomaly. To derive boundary Weyl anomaly of $O(\epsilon^3)$, one can
work out bulk solutions (\ref{bulkmetric},\ref{bulkscalar}) of order
$O(\epsilon^3)$ and then select the UV logarithmic divergent terms in
the action. However, the $O(\epsilon^3)$ solutions are quite
complicated. Instead, we use a simpler method developed by
\cite{Dong:2016wcf}, which needs only bulk solutions of order
$O(\epsilon^2)$.

Consider the variation of the gravitational action (\ref{action}), we
have
\begin{eqnarray} \label{dI}
&&\delta I=\int_N \text{EOM}+\int_Q \sqrt{\gamma} \left[
    \left((K-T+\frac{\xi}{2} \hat{\phi})\gamma^{ij} -K^{ij}
    \right)\delta \gamma_{ij}+( \xi-\hat{n}^{\mu}\hat{\nabla}_{\mu}
    \hat{\phi})\delta \hat{\phi} \right]\nonumber\\ &&\ \ \ \ \ +
  \int_{M} \sqrt{g} \left( \frac{1}{2} T^{ij}_{\text{non-ren}} \delta
  g_{ij} + O_{\text{non-ren}} \delta \phi \right),
\end{eqnarray}
where the first line of (\ref{dI}) vanishes due to EOM and NBC
(\ref{NBCmetric},\ref{NBCscalar}), $T^{ij}_{\text{non-ren}}$ and
$O_{\text{non-ren}}$ are non-renormalized stress tensor and one point
function of scalar, respectively.  To get renormalized stress tensor
and scalar operator, we can subtract a reference one without
boundaries. For the reference action without bulk boundary $Q$, we
have
\begin{eqnarray} \label{dI0}
\delta I_0= \int_{M} \sqrt{g} \left( \frac{1}{2} T^{ij}_{0} \delta
g_{ij} + O_{0} \delta \phi \right),
\end{eqnarray}
where the integration is over the same region $M$ as in
(\ref{dI}). Consider the difference of (\ref{dI}) and (\ref{dI0}), we
get
\begin{eqnarray} \label{dI-I0}
\delta (I-I_0)= \int_{M} \sqrt{g} \left( \frac{1}{2}
T^{ij}_{\text{holo}}\delta g_{ij} + O_{\text{holo}} \delta \phi
\right),
\end{eqnarray}
where $T^{ij}_{\text{holo}} :=T^{ij}_{\text{non-ren}} -T^{ij}_{0}$ is
the renormalized holographic stress tensor and similarly for
$O_{\text{holo}}$.  Select the UV logarithmic divergent term of above
equation and notice that $I$ and $I_0$ have the same bulk Weyl
anomaly, we obtain
\begin{eqnarray} \label{dAholo}
  \delta \mathcal{A}|_{\partial M}=\delta (I-I_0)_{\log \epsilon}=\int_{M} \sqrt{g}
  \left( \frac{1}{2}
T^{ij}_{\text{holo}} \delta g_{ij} + O_{\text{holo}} \delta \phi
\right) \Big|_{\log \epsilon},
\end{eqnarray}
which is just the holographic derivation  of (\ref{key}). The key point here
is that the left hand of (\ref{dAholo}) is a total variation.  As a
result, the boundary Weyl anomaly can be obtained by integrating
$\delta g_{ij}$ and $\delta \phi$.  Since we are interested in the scalar
contributions to Weyl anomaly, we can turn off the variation of
metric. By integrating (\ref{dAholo}), we can obtain Weyl anomaly up
to some irrelevant bulk terms such as $\mathcal{A}_2$ (\ref{A2}). Here
by `irrelevant terms', we mean `integration constant' terms which do not
contribute to $\delta \mathcal{A}|_{\partial M}$. (\ref{dAholo}) shows
that it is sufficient to derive $\delta \mathcal{A}|_{\partial M}$ of
$O(\epsilon^3)$ from $O_{\text{holo}}$ of $O(\epsilon^2)$, due to the
fact that $\phi$ is of $O(\epsilon)$.

Recall that, in section 4.1, we have obtained the holographic scalar
operator $O_{\text{holo}}$ as (\ref{FermicondensationGoodI}) with
boundary central charges given by (\ref{b3}, \ref{b4}, \ref{b1},
\ref{b2}). Substituting $O_{\text{holo}}$ into (\ref{dAholo}) and
integrating $\delta \phi$, we get the holographic boundary Weyl
anomaly as (\ref{anomaly}) with boundary central charges given by
(\ref{b3}, \ref{b4}, \ref{b1}, \ref{b2}).  This is just a turn-around
of the logic of section 3.1. Thus there is no need to repeat the
calculations here.  Note that, from (\ref{dAholo}) one cannot derive
all of the bulk Weyl anomaly.

\section{Holographic Story II: CFT without Boundary}

In this section, we give a holographic derivation of the anomalous
transformation rule (\ref{FermicondensationI3}) for the scalar operator $O$
under Weyl transformation. 

According to \cite{Imbimbo:1999bj} , the Weyl transformations
$g'_{ij}=e^{-2\sigma} g_{ij}$ can be realized by suitable bulk
diffeomorphisms. Inspired by \cite{Imbimbo:1999bj}, we take the ansatz
\cite{Zheng:2019xeu}
\begin{eqnarray}\label{diffeomorphisms1}
&&\rho=\rho' e^{2\sigma(x')} \left(1+\sum_{n=1}^{\infty} \rho'^n
  b_{(n)}(x') \right)\\ &&x^i=x'^i+\sum_{n=1}^{\infty} \rho'^n
  a^i_{(n)}(x') \label{diffeomorphisms2}
\end{eqnarray}
which is non-perturbative in the conformal factor.  We require that
the above diffeomorphisms leave the form of bulk metric
(\ref{AdSmetric}) invariant, i.e.,
\begin{eqnarray}\label{diffeomorphisms1ri}
&&G'_{\rho \rho}=\frac{\partial X^{\mu}}{\partial \rho'}
  \frac{\partial X^{\nu}}{\partial \rho'}
  G_{\mu\nu}=\frac{1}{4\rho'^2},\\ &&G'_{\rho i}=\frac{\partial
    X^{\mu}}{\partial \rho'} \frac{\partial X^{\nu}}{\partial x'^i}
  G_{\mu\nu}=0 .\label{diffeomorphisms2ri}
\end{eqnarray}
 Substituting (\ref{diffeomorphisms1},\ref{diffeomorphisms2}) into
 (\ref{diffeomorphisms2ri}), we obtain
 \cite{Imbimbo:1999bj,Zheng:2019xeu}
 \begin{eqnarray}\label{ai}
&&a_{(1)}^i=- \frac{1}{2} g'^{ij}\partial_j \sigma,\\ &&
   b_{(1)}=-\frac{1}{2} g'^{ij}\partial_i \sigma \partial_j
   \sigma, \label{bi}
\end{eqnarray}
where $g'{}_{}^{ij}=e^{2\sigma}g_{}^{ij}$ is non-perturbative in the
scale factor.

Now we are ready to derive the transformation law of scalar operator
$O$ under Weyl transformation. Under the diffeomorphisms
(\ref{diffeomorphisms1},\ref{diffeomorphisms2}), the bulk scalar
(\ref{AdSbdyscalar}) becomes
\begin{eqnarray}\label{dscalar}
&&\hat{\phi'}(\rho', x')=\hat{\phi}(\rho,
  x)=\rho^{\frac{1}{2}}\phi(x)+\rho^{\frac{3}{2}}[
    \phi_{(2)}(x)+\psi_{(2)}(x)\ln
    \rho]+O(\rho^{\frac{3}{2}})\nonumber\\
  &=&\rho'^{\frac{1}{2}}e^{\sigma}
  (1+\frac{1}{2}\rho'
  b_{(1)}) \left(\phi(x')+\rho' a_{(1)}^i\partial_i \phi(x') \right)
  \nonumber\\
  &&+ \rho'^{\frac{3}{2}}e^{3\sigma} (1+\frac{3}{2}\rho'
  b_{(1)}) \Big( \phi_{(2)}(x')+\psi_{(2)}(x')\ln \rho' +2\sigma
  \psi_{(2)}(x') \Big) +O(\rho'^{\frac{3}{2}})
  \nonumber\\
  &=&\rho'^{\frac{1}{2}} [ e^{\sigma} \phi(x')]
  +O(\rho'^{\frac{3}{2}}) \nonumber\\
  &&+ \rho'^{\frac{3}{2}} \left[
    e^{\sigma} \left(a_{(1)}^i\partial_i
    \phi(x')+\frac{1}{2}b_{(1)}\phi(x') \right) + e^{3\sigma} \Big(
    \phi_{(2)}(x')+2\sigma \psi_{(2)}(x')+\psi_{(2)}(x')\ln \rho'
    \Big) \right].\nonumber\\
\end{eqnarray}
From the above equation and (\ref{AdSbdyscalar}) , we can read off the
transformation rules
\begin{eqnarray}\label{tranrule1}
&&\phi'=e^{\sigma} \phi ,\\ \label{tranrule2}
  &&\psi'_{(2)}=e^{3\sigma} \psi_{(2)},\\ \label{tranrule3}
&&\phi'_{(2)}=e^{3\sigma} \phi_{(2)} + e^{\sigma} \left(a_{(1)}^i\partial_i
    \phi(x')+\frac{1}{2}b_{(1)}\phi(x') \right) + e^{3\sigma} \Big(
  2\sigma \psi_{(2)}(x')
    \Big)
\end{eqnarray}
According to the standard approach, $\psi_{(2)}$ can be obtained from
either EOM (\ref{EOMscalar}) or the variation of holographic Weyl
anomaly (\ref{bulkanomaly}). Applying both methods, we get
\begin{eqnarray}\label{Janomaly1}
&& \psi_{(2)}=-\frac{1}{4} \nabla^2 \phi+\frac{1}{24} R \phi+\frac{1}{12}\phi^3, \\ &&
  \psi'_{(2)}=-\frac{1}{4} \nabla'^2 \phi'+\frac{1}{24} R'\phi'+\frac{1}{12}\phi'^3.
  \label{Janomaly2}
\end{eqnarray}
One can check that (\ref{Janomaly1}) and (\ref{Janomaly2}) obey the
transformation rule (\ref{tranrule2}), which is a test of our
results. Substituting (\ref{ai},\ref{bi},\ref{Janomaly2}) into
(\ref{tranrule3}) and noting that $\la O\ra=2\phi_{(2)}$, we finally obtain
the
Weyl transformation rule
\begin{eqnarray}\label{holscalarlaw}
  \la O\ra =e^{-3\sigma} \la O \ra' +
  \nabla(\sigma\nabla\phi)-\frac{1}{6}\phi R \sigma
+\frac{1}{2}\phi (\nabla\sigma)^2
-\frac{1}{3}\phi^3 \sigma,
\end{eqnarray}
in signature $(-1,1,1,1)$ or $(1,1,1,1)$. Transforms into signature
$(1,-1,-1,-1)$, $\nabla(\sigma\nabla\phi), (\nabla\sigma)^2$ change
sign and (\ref{holscalarlaw}) agrees with the field-theoretical result
(\ref{FermicondensationI3}) with central charges
(\ref{bulkcentralcharge}).

\section{Conclusions and Discussions}

In this paper, we have investigated anomalous Fermi condensation (one
point function of scalar operator) due to Weyl anomaly. We obtain
general form of Weyl anomaly due to a background scalar for 4d BCFTs, 
which consequently leads to
two kinds of anomalous Fermi condensation. The first kind
occurs near a boundary, while the second kind appears in conformally
flat spacetime without boundaries. It is interesting that the first
kind of Fermi condensation could be non-zero in flat spacetime and
even if there is no background scalar. While the second kind of Fermi
condensation only appears in a curved spacetime with non-zero
background scalar. We verify our results with free BCFT and
holographic BCFT. In particular, we consider carefully the back
reaction to the AdS geometry due to the scalar field and reproduce
precisely the  shape and curvature dependence of the field theoretic
Fermi condensate from the holographic one
point function.

For simplicity, we focus on CFT/BCFTs in four
dimensions in this paper. It is interesting to generalize our works to
general dimensions. Besides, it is also interesting to study Fermi
condensation for general QFT. For QFT, more possible terms are allowed
in Weyl anomaly, which would correct the results of anomalous Fermi
condensation. We hope to address these problems in future.

\section*{Acknowledgements}

We would like to thank Ting-Wai Chiu, Bei-Lok Hu, Satoshi Iso, Gary
Shiu, L. Shu, X. Gao and Y. Zhou for useful discussions and
comments. R. X. Miao thank the hospitality during the workshops
``Boundaries and Defects in Quantum Field Theory'' and ``East Asia
Joint Workshop on Fields and Strings 2019'', where parts of the work
are worked out.  C. S. Chu is supported by the MOST grant
107-2119-M-007-014-MY3. R. X. Miao acknowledges the supports from NSFC
grant (No. 11905297) and Guangdong Basic and Applied Basic Research
Foundation (No.2020A1515010900).

\appendix

\section{Weyl Anomaly in Different Signatures} 

In this appendix, we clarify that Weyl anomaly $<T^i_{\ i}>$ differs
by a minus sign in different signature. First, let us stress that the
action is independent of the choice of signature.  In signature
$(-1,1,1,1)$, the stress tensor is defined by \cite{Carroll:2004st}
  \begin{eqnarray} \label{TijsignatureI}
 \delta I =\frac{1}{2} \int_M \sqrt{|g|} \hat{T}^{ij}\delta \hat{g}_{ij},
 \end{eqnarray}
while in signature $(1,-1,-1,-1)$ the stress tensor is defined by
\cite{Parker:2009uva}
  \begin{eqnarray} \label{TijsignatureII}
 \delta I =-\frac{1}{2} \int_M \sqrt{|\hat{g}|} T^{ij}\delta
 g_{ij} ,
 \end{eqnarray}
where $\hat{g}_{ij}=-g_{ij}$. From (\ref{TijsignatureI}) and
(\ref{TijsignatureII}), we notice that $T^{ij}=\hat{T}^{ij}$ and hence
the Weyl anomaly in different signature differs by a minus sign
  \begin{eqnarray} \label{AIandAII}
\la T^{ij} \ra g_{ij} =- \la\hat{T}^{ij} \ra \hat{g}_{ij}.
 \end{eqnarray}
  The Euclidean theory is related to the theory with signature
  $(-1,1,1,1)$ by a Wick rotation, therefore the Weyl anomaly in Euclidean theory is
  also different by a minus sign from the Weyl anomaly in the signature
  $(1,-1,-1,-1)$.

For the convenience of readers, let us list some important formulas in
both signature.  The action of Dirac field takes the form
\begin{eqnarray} \label{Fermiactionappecdix}
I=\int_M \sqrt{|g|} \left( \bar{\psi}i \gamma^{i}\nabla_{i}\psi+ \phi
\bar{\psi} \psi\right),
\end{eqnarray}
where $\bar{\psi}=\psi^+ \gamma^0$, $(\gamma^0)^+=\gamma^0$,
$(\gamma^a)^+=-\gamma^a$ and the gamma matrix obeys
\begin{eqnarray}
\{ \gamma^{i}, \gamma^{j} \}=2\eta \ g^{ij}.
\end{eqnarray}
Here $\eta=-1$ for signature $(-1,1,1,1)$ \cite{Carroll:2004st} and
$\eta=1$ for signature $(1,-1,-1,-1)$ \cite{Parker:2009uva}. The key
relation (\ref{key}) becomes
\begin{eqnarray} \label{key-2}
\eta (\delta_{\phi} \mathcal{A})_{\partial M}=\left(\int_M \sqrt{|g|}
\la \bar{\psi}\psi \ra \delta \phi \right)_{\log\epsilon}.
\end{eqnarray}

To summarize,
the action, the stress tension $T_{ij}$, the gamma
matrices $\gamma^i$ and the Fermi condensation $\la\bar{\psi}\psi\ra $ are the
same, while the metric $g_{ij}$ and the Weyl anomaly $\la T^i_{\ i}\ra$ differ
by a minus sign in different signatures.  
Note that we take signature $(1,-1,-1,-1)$ from section 1 to section
4, while signature $(-1,1,1,1)$ or $(1,1,1,1)$ in section 5 and
section 6 in this paper.  To avoid confusion, we denote Weyl anomaly
in signature $(1,-1,-1,-1)$ by $\mathcal{A}$ and Weyl anomaly in
signature $(-1,1,1,1)$ by $\mathcal{\hat{A}}$ in the main text of this
paper.

\section{Solutions in the Flat Limit}

In the flat limit, the bulk metric is given by (\ref{bulkmetricflatlimit})
with
  \begin{eqnarray} \label{X1e}
  X_{1e}(s)=d_1 \left(\sqrt{s^2+1} \left(d_3-\phi
  _0\right)+\frac{d_3-\phi _0}{\sqrt{s^2+1}}-\frac{d_3}{2
    \left(s^2+1\right)}-\frac{1}{2} d_3 \log
  \left(s^2+1\right)-\frac{3 d_3}{2}+2 \phi _0\right), 
 \end{eqnarray}
   \begin{eqnarray} \label{X2}
  X_{2}(s)=-\frac{d_3}{2} \left(-s^2+2 \sqrt{s^2+1}-2\right) \phi
  _0+d_3^2 \left(-\frac{s^2}{4}+\sqrt{s^2+1}-\frac{1}{4} \log
  \left(s^2+1\right)-1\right)-\frac{1}{12} s^2 \phi _0^2,\;\;\;\;
 \end{eqnarray}
    \begin{eqnarray} \label{X2e}
  X_{2e}(s)=\frac{d_1}{2} \left[d_5 \Big(\frac{4 \left(s^2+2\right)
    (\sqrt{s^2+1}-1)}{\sqrt{s^2+1}}-\left(s^2+4\right) \log
  \left(s^2+1\right)\Big)+\phi _1 \left(\left(s^2+4\right) \log
  \left(s^2+1\right)-4 s^2\right)\right],\;\;\;\;
 \end{eqnarray}
     \begin{eqnarray} \label{g1e}
  g_{1e}(s)=\frac{d_1 \left(4 \left(s^2+1\right)^{3/2} \left(d_3-\phi
    _0\right)+4 \left(s^2+1\right)^{5/2} \left(d_3-\phi _0\right)+2
    d_3 \left(s^2+1\right)-d_3\right)}{12 \left(s^2+1\right)^2}+d_1
  \left(\frac{2 \phi _0}{3}-\frac{3 d_3}{4}\right),\;\;\;\;
 \end{eqnarray}
    \begin{eqnarray} \label{g2}
  g_{2}(s)=\frac{1}{24} \Big(-4 d_3 (-s^2+2
  \sqrt{s^2+1}-2 ) \phi _0+\frac{d_3^2 \left(-3 s^4+ 4 (2
    \sqrt{s^2+1}-3 ) s^2+8
    (\sqrt{s^2+1}-1)\right)}{s^2+1}-2 s^2 \phi _0^2\Big),\;\;
   \end{eqnarray}
\begin{eqnarray} \label{g2e}
  g_{2e}(s)=
  \frac{
    d_1 \Big(
    d_5-\phi_1 \left(s^2\left(s^2+2\right)
    -2 \left(s^2+1\right) \log\left(s^2+1\right)
    \right)
    \Big)
  }
  {4 \left(s^2+1\right)}.
\end{eqnarray}
The embedding function of $Q$ takes the form (\ref{bulkQ})
\begin{eqnarray}\label{bulkQflatlimit}
x=-\sinh \rho z +\epsilon \xi \lambda_{1e} z^2 +\epsilon^2
(\lambda_2+\xi \lambda_{2e}) z^3+ \cdots
\end{eqnarray}
with
 \begin{eqnarray} \label{lambda1}
   \lambda_{1e}&=&\frac{\phi _0 \cosh ^3\rho }{288 (7 \sinh \rho
    -\sinh (3 \rho )+\cosh \rho -\cosh (3 \rho ))}
  \Big[16 \cosh \rho -96 \cosh (3 \rho )+16 \cosh (5 \rho ) \nn\\
    &&   +\sinh\rho (-92+6 \log (\coth ^2\rho ))
    +\sinh (3 \rho )(-84 +3  \log (\coth ^2\rho)) \nn\\
    && 
    +\sinh (5 \rho )(16 -3 \log (\coth ^2\rho))
     \Big],
 \end{eqnarray}
  \begin{eqnarray} \label{lambda2}
  \lambda_{2}&=&\frac{\phi _0^2 \sinh \rho}{2304 (\cosh \rho -2
    \sinh \rho )^2 (\sinh \rho +\cosh \rho )^2}
  \Big[
    12 +28 \sinh (2 \rho )+112 \sinh (4 \rho)-20 \sinh (6 \rho )
    \nonumber\\
  &&
    +\cosh (2 \rho )\left(4-3 \log(\coth ^2\rho)\right)
    +\cosh (4 \rho ) \left(100-6 \log (\coth^2\rho )\right)\nn\\
  &&  -\cosh(6 \rho )(20 -3  \log(\coth ^2\rho))
  + 6 \log \left(\coth ^2\rho \right)
    \Big],\;\;
 \end{eqnarray}
 \begin{eqnarray} \label{lambda2e}
  \lambda_{2e}&=&\frac{\phi _1 \cosh ^4(\rho )}{72 (\sinh (2 \rho
    )+\cosh (2 \rho )-3)}
  \Big[
    48-8\sinh (2 \rho )+4 \sinh (4 \rho ) +3 \log (\coth ^2\rho )
    \nonumber\\
    &&
    -2 \cosh (2 \rho) \left(9 \log (\coth ^2\rho)+10\right)
    +\cosh (4 \rho ) \left(3 \log (\coth ^2\rho)+4\right)
    \Big].
 \end{eqnarray}

 \section{Back Reaction due
   to Scalar BC}

In the main text of the paper, we focus on solutions in the linear order of
$\xi$, where $\xi$ labels the NBC (\ref{NBCscalar})
of the scalar field. In this appendix, we discuss solutions in higher
orders of $\xi$ briefly.  For simplicity, we focus on both the
flat limit with $k_{ab}=q_{ab}=0$ and the no-scalar limit
$\phi=0$. Then the ansatz for bulk metric, bulk scalar and embedding
function of $Q$ become
 \begin{eqnarray}\label{bulkmetrice2}
 ds^2=\frac{1}{z^2}\Big{[} dz^2+ \left(1+\xi^2
   X_{e}(\frac{z}{x})\right)dx^2 +\delta_{ab} \left(1+\xi^2
   \ g_{e}(\frac{z}{x}) \right)dy^a dy^b\Big{]} + O(\xi^3,
 \epsilon)
\end{eqnarray}
\begin{eqnarray}\label{bulkscalare2}
\hat{\phi}=f_0(\frac{z}{x})+ \xi^3\ f_{e}(\frac{z}{x})
+O(\xi^4, \epsilon),
\end{eqnarray}
and
\begin{eqnarray}\label{bulkQe2}
x=-\sinh \rho \ z +\xi^2 \lambda_0 z+ +O(\xi^3,
\epsilon),
\end{eqnarray}
where $f_0$ (\ref{f0Sol},\ref{f0Sol1}) is of order $\xi$.
Following approach of section 4.1, we can solve the coupled
Einstein-scalar EOM (\ref{EOMmetric},\ref{EOMscalar}) with DBC
(\ref{fXQ}) on $M$ and NBC (\ref{NBCmetric},\ref{NBCscalar}) on $Q$.
We obtain
\begin{eqnarray}\label{SolutionsXee2}
  &&X_e(s)=-\frac{d_1^2 \left(2 \left(s^2+1\right)^2
    \log (s^2+1)+(2 s^4-3 s^2-2) s^2\right)}{32
    (s^2+1)^2}\\ \label{Solutionsgee2}
  &&g_e(s)=\frac{1}{32} d_1^2 \left(\frac{s^2 (s^4+5
    s^2+2)}{\left(s^2+1\right)^3}-2 \log
  (s^2+1)\right)\\ \label{Solutionsfee2}
  &&f_e(s)=\frac{s^3
    \left[
      d_1^3 \left(-15 s^4-12 s^2+6 (s^2+1)^2 \log
      (s^2+1)-2\right)+64 (s^2+1)^3 v_1\right]}
         {64 \left(s^2+1\right)^{9/2}}\\ \label{Solutionsl0e2}
  &&\lambda_0=\frac{\cosh \rho  \coth ^3\rho  \left(-30 \cosh (2
    \rho )+\cosh (4 \rho )-8 \sinh ^2\rho  \cosh ^4\rho  \log
    \left(\coth ^2\rho \right)+37\right)}{2304}\nonumber\\
\end{eqnarray}
where $d_1$ is given by (\ref{f0Sol1}) and $v_1$ is given by
\begin{eqnarray}\label{v1e2}
v_1=-\frac{\cosh ^3\rho  \left(399 \cosh (2 \rho )+6 \cosh (4 \rho
  )+\cosh (6 \rho )-886\right)  \coth ^3\rho
}{27648}.
\end{eqnarray}
Substituting (\ref{f0Sol},\ref{bulkscalare2},\ref{Solutionsfee2}) into
(\ref{scalarcurrent}), we obtain
\begin{eqnarray} \label{O0-higher}
\la O \ra=\frac{2 d_1 \xi+(2
  v_1-\frac{d_1^3}{16})\xi^3}{x^3}+O(1/x^2, \xi^4).
\end{eqnarray}
which gives the central charge
\begin{eqnarray} \label{b3e2}
b_3&=&\frac{1}{6} \left( d_1 \xi+(
v_1-\frac{d_1^3}{32})\xi^3\right)+O(\xi^4)\nonumber\\
&=&
-\frac{\xi}{18}  \cosh ^3(\rho ) \coth \rho
- \frac{\xi^3 \cosh^3 \rho \left(3 \cosh (2 \rho)-7 \right)
  \coth ^3\rho}{1296}+O\left(\xi^4\right).
\end{eqnarray}

\end{document}